\documentclass[10pt,twocolumn,showpacs,amsmath,amssymb,floatfix,superscriptaddress]{revtex4-2}

\usepackage{graphicx}% Include figure files
\usepackage{dcolumn}% Align table columns on decimal point
\usepackage{bm}
\usepackage{color}
\usepackage{txfonts}
\usepackage{microtype}
\usepackage{hyperref}
\usepackage[english]{babel}
\usepackage{slashed}
\usepackage{gensymb}
\usepackage{epsfig}
\usepackage[normalem]{ulem}
\usepackage{amsmath}
\usepackage{array}
\usepackage{booktabs}

\usepackage{CJKutf8}

\allowdisplaybreaks[4]

\begin{document}

\renewcommand{\figurename}{FIG}	
\bibliographystyle{apsrev4-2}
	
\title{$\beta$-Decay Half-Lives Serve as Novel Evidence for the New Magic Number  \(N=32\) }

\author{L. Guo (\begin{CJK}{UTF8}{gkai}郭亮\end{CJK})}
\affiliation{School of Nuclear Science and Technology, Lanzhou University, Lanzhou 730000, China}
\affiliation{Frontiers Science Center for Rare isotopes, Lanzhou University, Lanzhou 730000, China}

\author{Z. H. Wang (\begin{CJK}{UTF8}{gkai}王之恒\end{CJK})}
\affiliation{School of Nuclear Science and Technology, Lanzhou University, Lanzhou 730000, China}
\affiliation{Frontiers Science Center for Rare isotopes, Lanzhou University, Lanzhou 730000, China}

\author{X. L. Zhi (\begin{CJK}{UTF8}{gkai}植学林\end{CJK})}
\affiliation{School of Nuclear Science and Technology, Lanzhou University, Lanzhou 730000, China}
\affiliation{Frontiers Science Center for Rare isotopes, Lanzhou University, Lanzhou 730000, China}

\author{Y. F. Niu (\begin{CJK}{UTF8}{gkai}牛一斐\end{CJK})}\email{niuyf@lzu.edu.cn}
\affiliation{School of Nuclear Science and Technology, Lanzhou University, Lanzhou 730000, China}
\affiliation{Frontiers Science Center for Rare isotopes, Lanzhou University, Lanzhou 730000, China}
\affiliation{Institute of Modern Physics, Chinese Academy of Sciences, Lanzhou 730000, China}
\affiliation{Department of Nuclear Physics, China Institute of Atomic Energy, Beijing, 102413, China}

\author{W. H. Long (\begin{CJK}{UTF8}{gkai}龙文辉\end{CJK})}
\affiliation{School of Nuclear Science and Technology, Lanzhou University, Lanzhou 730000, China}
\affiliation{Frontiers Science Center for Rare isotopes, Lanzhou University, Lanzhou 730000, China}

\author{Z. M. Niu (\begin{CJK}{UTF8}{gkai}牛中明\end{CJK})}
\affiliation{School of Physics and Optoelectronic Engineering, Anhui University, Hefei 230601, China}
%\affiliation{Institute of Physical Science and Information Technology, Anhui University, Hefei 230601, China}

\author{Q. B. Zeng (\begin{CJK}{UTF8}{gkai}曾全波\end{CJK})}
\affiliation{Institute of Modern Physics, Chinese Academy of Sciences, Lanzhou 730000, China}
\affiliation{School of Nuclear Science and Technology, University of Chinese Academy of Sciences, Beijing 100049, China}

\author{Z.  Liu (\begin{CJK}{UTF8}{gkai}刘忠\end{CJK})}
\affiliation{Institute of Modern Physics, Chinese Academy of Sciences, Lanzhou 730000, China}
\affiliation{School of Nuclear Science and Technology, University of Chinese Academy of Sciences, Beijing 100049, China}

	\date{\today}% It is always \today, today,
	%  but any date may be explicitly specified

\begin{abstract}

Conventional signatures of nuclear magic number, including low-lying quadrupole collectivity and mass systematics, face significant challenges when probing emergent shell closures near the drip line. 
However, $\beta$-decay half-lives are among the first experimental observables measurable following the discovery of neutron-rich isotopes. 
This letter demonstrates that $\beta$-decay half-lives provide evidence for the emergent magic number 
$N=32$. 
The observed half-life pattern around the $N=32$ can be attributed to the occupation probabilities of orbitals above this shell gap, which directly reflect the gap's magnitude. 
Our results reveal a pronounced $N=32$ shell gap in Ca isotopes and a weaker yet apparent gap in K isotopes, consistent with mass and electromagnetic transition data. Furthermore, the analysis indicates no prominent closed-shell signature at $N=32$ in Ar and Cl isotopes.

%Conventional observables including low-lying quadrupole states and nuclear mass face significant challenges in exploring the emerging magic numbers in exotic nuclei near the drip-line. 
%In contrast, the measurements on $\beta$-decay half-life in exotic nuclei are more accessible. 
%In this letter, we demonstrate that $\beta$-decay half-lives provide evidence for the new magic number  \(N=32\).  It is shown that
%the correlation between \(\beta\)-decay half-lives and the \(N=32\) shell structure is intimately tied to the occupation probabilities of the orbitals above $N=32$ shell, which reflect the magnitude of the shell gap. 
%Our findings reveal a strong \(N=32\) shell gap in Ca isotopes and a diminishing one in K isotopes, in agreement with mass and electromagnetic transition data. 
%Moreover, our analysis indicates that the closed-shell signature at \(N=32\) is not prominent in Ar and Cl isotopes. 
%This approach provides a refined perspective on probing new magic numbers through nuclear weak interactions.
\end{abstract}
	
\maketitle

Magic numbers represent fundamental characteristics of atomic nuclei, distinct from those of atoms owing to strong spin-orbit coupling  \cite{Mayer1948,Haxel1949}.
The development of radioactive ion beam facilities has significantly broadened nuclear physics from stable nuclei to exotic nuclei far from stability \cite{Otsuka2001, Meng2006,Sorlin2008}.
In this context, the nuclear structure undergoes distinct—and sometimes dramatic—changes, characterized by the vanishing of traditional magic numbers such as $N = 8$, 20, and 28 in neutron-rich $^{11}$Li \cite{Simon1999}, $^{32}$Mg \cite{Motobayashi1995} and $^{42}$S \cite{Bastin2007}, alongside the emergence of new magic numbers\cite{Ozawa2000,Stanoiu2004, Doornenbal2007,Hoffman2008,Kanungo2009,Tshoo2012}, as reviewed in  Refs. \cite{Tanihata2013,Otsuka2020}.

Special attentions are paid to the occurrence of new magic numbers $N = 32$ in neutron-rich $pf$-shell nuclei recently.
Experimentally, the magic nature at $N = 32$ has been  prominently observed in Ca \cite{Huck1985, Gade2006}, Ti \cite{Janssens2002,Dinca2005}, and Cr \cite{Prisciandaro2001,Buerger2005} isotopes, as evidenced by the enhanced excitation energy of the $2_{1}^{+}$ state, $E(2_{1}^{+})$, and the reduced transition strength $B(E2)$.
High-precision mass measurements of K \cite{Rosenbusch2015}, Ca \cite{Wienholtz2013}, and Sc \cite{Xu2015,Xu2019}, along with the one-neutron knockout experiment  in $^{52}$Ca \cite{Enciu2022} further confirm the shell closure at $N = 32$.
However, the robustness of this magic number remains unclear due to inconsistent experimental evidences.
%However, several ambiguities remain regarding this magic number. 
These include the unexpectedly large increases in charge radii of Ca \cite{GarciaRuiz2016} and K \cite{Koszorus2021} isotopes and  the presence of weak shell effects at $N = 32$ through mass measurements in Ti isotopes \cite{Leistenschneider2018}. 
In addition, for argon isotopes, the current evidence for $N=32$ magicity rests solely on the $E(2_{1}^+)$ measurement in $^{50}$Ar \cite{Steppenbeck2015,Cortes2020}, however, the modest energy increase renders the evidence inconclusive \cite{Sahoo2025}. 
%leaves the shell closure interpretation ambiguous. 
% and the low $E(2_{1}^+)$ observed in $^{50}$Ar  \cite{Steppenbeck2015,Cortes2020}.
%In addition, for Ar isotopes,  the $E(2_{1}^+)$ in $^{50}$Ar  \cite{Steppenbeck2015,Cortes2020}  is so far the only measurement for the hint of $N=32$'s magic nature, however, ambiguity exists due to the less significant increase of  $E(2_{1}^+)$ at $N=32$. 

%has led to divergent interpretations of  $N=32's$ magic nature in argon isotopes \cite{}.
%For shell closure $N = 34$, it has been suggested in $^{54}$Ca through enhanced $E(2_{1}^+)$ \cite{Steppenbeck2013} and mass measurement \cite{Michimasa2018}, as well as in $^{52}$Ar through enhanced $E(2_{1}^+)$ \cite{Liu2019}. 
%For nuclei with $Z > 20$,  both the low-lying quadrupole state \cite{Liddick2004,Dinca2005,Steppenbeck2017} and mass \cite{Leistenschneider2021,Iimura2023} measurements provide no evidence for the shell closure at $N = 34$.

Theoretically, the closed-shell structure at \(N = 32\) in Ca isotopes can be reproduced using relativistic Hartree-Fock-Bogoliubov (RHFB) theory with the PKA1 interaction \cite{Li2016,Liu2020}. 
Skyrme Hartree-Fock-Bogoliubov (SHFB) theory shows that incorporating the tensor force enhances the magicity at this neutron number in Ca isotopes\cite{Grasso2014}. 
Moreover, the low-lying quadrupole states have been examined to further investigate the magic nature at \(N = 32\)  through various approaches, including large-scale shell model calculations \cite{Honma2002,Janssens2002,Mantica2003,Liddick2004,Honma2005,Coraggio2009}, the variation after projection generator coordinate method \cite{Rodriguez2007}, the coupled-cluster method with chiral effective nuclear force \cite{Hagen2012}, in-medium similarity renormalization group method \cite{Sahoo2025},
and the configuration-interaction relativistic Hartree-Fock (RHF) model \cite{Liu2024} with PKA1 interaction.

Due to the low production yields of these neutron-rich isotopes, experimental measurements  on  low-lying quadrupole states and nuclear masses are challenging.  
In contrast, the measurements on $\beta$-decay half-life in exotic nuclei are more accessible. 
However, the relationship between $\beta$-decay half-lives and magic numbers is complex. For example, different  $\beta$-decay half-life patterns  have been observed around  traditional magic numbers \cite{Zhang2007, Sorlin1993}, while conversely, similar patterns may also arise from nuclear shape transitions \cite{Yoshida2023}.  %This complexity arises because both $Q_\beta$ values and transition strengths play crucial roles, and these depend critically on factors like shell gaps, deformation, and specific shell orbitals.  
Therefore, careful analysis is needed to identify magic numbers from $\beta$-decay half-lives. 
Systematic measurements of $\beta$-decay half-lives for twenty neutron-rich nuclei around $^{78}$Ni reveal that the pronounced $N = 50$ shell gap enhances the $\beta$-decay energy through first-forbidden (FF) transitions, resulting in a sudden shortening of half-lives for Ni isotopes beyond $N = 50$ shell closure\cite{Yoshida2019, Xu2014}.  
%By observing  experimental $\beta$-decay half-lives near traditional magic numbers,  the correlation between shell closures and $\beta$-decay half-lives was discussed \cite{Zhang2007, Sorlin1993}. 
%the $\beta$-decay half-life—reflecting the weak interaction in nuclei—provides a distinct sensitivity to shell gaps \cite{Sorlin1993,Sorlin1998,Zhang2007,Daugas2011} and is especially valuable for exotic nuclei, where such measurements are more accessible.
%Moreover, Ref. \cite{Sorlin1998} points out that the quick drop of the half-life observed at $N = 33$ for $^{53}$Ca is associated with the shell gap at $N = 32$. 
%However, a clear correlation between $\beta$-decay half-lives and new magic numbers has not been discussed. Once this correlation and its behind mechanism are revealed, $\beta$-decay half-lives can serve as a new evidence for the magic nature of $N=32$ in pf-shell nuclei. 
%despite these compelling correlations, the microscopic mechanism connecting $\beta$-decay half-lives to shell closure at $N = 32$ remains inadequately understood.
%Elucidating this mechanism is essential for gaining deeper insight into the interplay between weak interactions and nuclear shell evolution in exotic systems.
%By analyzing experimental $\beta$-decay half-lives around traditional magic numbers, \blue{different  $\beta$-decay half-life patterns   have been observed around the shell closures} \cite{Zhang2007, Sorlin1993}.
 However, the potential correlation between $\beta$-decay half-lives and new magic numbers (e.g., $N=32$ in $pf$-shell nuclei) remains unclear. 
While Ref. \cite{QBZeng2025} used newly measured half-lives to investigate such emergent shell closures, the suggestion of an $N=32$ shell closure in Cl isotopes stems primarily from shell model calculations. 
Crucially, understanding the observed $\beta$-decay pattern actually necessitates a weakened $N=32$  shell effect  through the configuration mixing.  
 Consequently, this shell closure finds little direct support in the experimental $\beta$-decay half-life pattern.
 Given that emergent shell closures are typically weaker than traditional magic numbers, it remains uncertain whether a direct correlation exists between $\beta$-decay half-life patterns and these new shell closures. 
 Therefore, investigating such a correlation is essential. 
 Once this correlation is established, along with its underlying mechanism, $\beta$-decay half-life measurements could potentially offer an additional signature for testing the magicity of $N=32$, complementing conventional shell-closure indicators.

In this letter, we investigate the correlation between  $\beta$-decay half-lives and nuclear shell structure at  $N = 32$  across Ca, K, Ar and Cl isotopes, examining the underlying physical mechanisms driving these phenomena. Self-consistent proton-neutron quasi-particle random phase approximation (pnQRPA) model is one of the most widely used microscopic models for the calculation of $\beta$-decay half-lives, which describes nuclei in the whole nuclear chart with the exception of few light ones employing a single energy density functional (EDF) \cite{Engel1999,Mustonen2016,Minato2022,Niksic2005,Niu2013,Marketin2016}.
Here, we adopt the PKA1 EDF  \cite{Long2007}, which explicitly includes the tensor coupling. This interaction successfully reproduces the emergence of new magic numbers of $N=32,34$ in calcium isotopes \cite{Li2016,Liu2020}. To incorporate PKA1 consistently, we have newly developed a  fully self-consistent pnQRPA model with tensor coupling based on RHFB theory \cite{Long2010,Geng2022,Geng2024},  providing new capabilities to probe $N=32$ shell-closure effects in $\beta$-decay half-lives. 
%with which the closed shell effects at $N=32$ on $\beta$-decay half-lives can be studied.
%extending its predictive power to  $\beta$-decay signatures near $N=32$. 

The details of the pnQRPA formalism in the canonical basis can be found in Refs. \cite{Paar2004,Niu2017}.
In the RHFB+ pnQRPA model both the direct and exchange terms are taken into account. 
Due to the presence of exchange terms, particle-hole ($p-h$) residual interaction contains terms generated not only by the isovector meson exchange ($\rho$ and $\pi$ ), but also by the exchange of isoscalar mesons ($\sigma$ and $\omega$). 
For PKA1 interaction, additionally, $\rho$-tensor coupling and $\rho$-vector-tensor coupling appear in the $p-h$ residual interaction with more details  in Ref. \cite{Wang2020}.
\iffalse
To remove the contact part of the $\pi$-pseudovector and $\rho$-tensor coupling, the corresponding zero-range counter terms are introduced, that is,
\begin{subequations}
	\begin{align}
		\hat{V}_{\pi-\mathrm{PV}}^\delta(1,2)=&-\frac{1}{3}\left[\frac{f_\pi}{m_\pi} \vec{\tau} \gamma_0 \gamma_5 \gamma^i\right]_1 \cdot\left[\frac{f_\pi}{m_\pi} \vec{\tau} \gamma_0 \gamma_5 \gamma_i\right]_2 \delta\left(\boldsymbol{r}_1-\boldsymbol{r}_2\right),\\
		%%%%%%%%%%%%%%%%%%%%%%%%%%%%%%%
		%%%%%%%%%%%%%%%%%%%%%%%%%%%%%%%
		\hat{V}_{\rho-\mathrm{T}}^\delta(1,2)&=\frac{1}{12 M^2}\left[f_\rho \gamma_0 \sigma_{\nu i} \vec{\tau}\right]_1 \cdot\left[f_\rho \gamma_0 \sigma^{\nu i} \vec{\tau}\right]_2 \delta\left(\boldsymbol{r}_1-\boldsymbol{r}_2\right),
	\end{align}
\end{subequations}
where $M$ and $m_\pi$ denote the nucleon and pion masses, respectively, and $f_\pi$, $f_\rho$ are the corresponding coupling constants. 
\fi
For the particle–particle ($p$–$p$) channel, both isovector and isoscalar pairing interactions are considered. 
The Gogny force D1S \cite{Berger1984} is used in the isovector $p$–$p$ channel for both the RHFB and pnQRPA calculations. 
For the isoscalar proton–neutron pairing in the pnQRPA, we adopt an Gogny-like interaction with $V_{0}$ denoting the overall strength of the isoscalar proton–neutron pairing \cite{Engel1999,Marketin2007}.
%\begin{equation}
%	\begin{aligned}
%		V_{T=0}(1,2)=-V_0 \sum_{j=1}^2 g_j e^{-\left[\left(\boldsymbol{r}_1-\boldsymbol{r}_2\right) / \mu_j\right]^2} \hat{\prod}_{S=1, T=0},
%	\end{aligned}
%\end{equation}
%with $\mu_{1}$ = 1.2 fm, $\mu_{2}$ = 0.7 fm, $g_{1}$ = 1, $g_{2}=-2$.
%The operator $\hat{\prod}_{S=1, T=0}$ projects onto states with spin $S = 1$ and isospin $T = 0$. 

Using the Gamow-Teller (GT) transition strengths $B_{m}$ for each excited state $|m\rangle$ obtained from QRPA calculations, we can calculate the $\beta$-decay half-life in the allowed GT approximation with the following formula,
\begin{equation}
	\begin{aligned}
		T_{1 / 2} & =\frac{D}{ g_A^2 \sum_m  {B_{m}} f(Z, E_m)},\label{Eq3}
	\end{aligned}
\end{equation}
where D = 6163.4 s and $g_{A}$ = 1. 
The summation is performed over all final states with excitation energies below the $Q_{\beta}$ value. 
The function $f(Z, E_m)$ represents the integrated phase volume, where $E_{m}$ is the $\beta$-decay transition energy. For more details, please see Ref. \cite{Niu2013}.
%The integrated phase volume $f(Z, E_m)$ is expressed as follows:
%\begin{equation}
%	\begin{aligned}
%		f(Z, E_m)=\frac{1}{m_{e}^{5}}\int_{m_{e}}^{E_{m}}p_e E_{e}\left(E_{m}-E_e\right)^2 F_{0}\left(Z, A, E_e\right)dE_{e},
%	\end{aligned}
%\end{equation}
%where $p_e$ , $E_e$, and $F_{0}(Z ,A, E_e)$ denote the momentum, energy, and Fermi function of the emitted electron, respectively \cite{Langanke2003}.
%The transition energy for $\beta$-decay, denoted as $E_m$, which reflects the energy difference between the initial and final states, is calculated using QRPA as follows:
%\begin{equation}
%	\begin{aligned}
%		E_m = \Delta_{np} - E_{\text{QRPA}},
%	\end{aligned}
%\end{equation}
%where $E_{\text{QRPA}}$ is the QRPA energy with respect to the ground state of the parent nucleus, corrected by the difference of the neutron and proton Fermi energies in the parent nucleus \cite{Engel1999}, and $\Delta_{np}$ represents the mass difference between neutron and proton.
\begin{figure}[!t]	
    \setlength{\abovecaptionskip}{0.cm}
	\setlength{\belowcaptionskip}{-0.cm}  	
	\centering\includegraphics[width=1\linewidth]{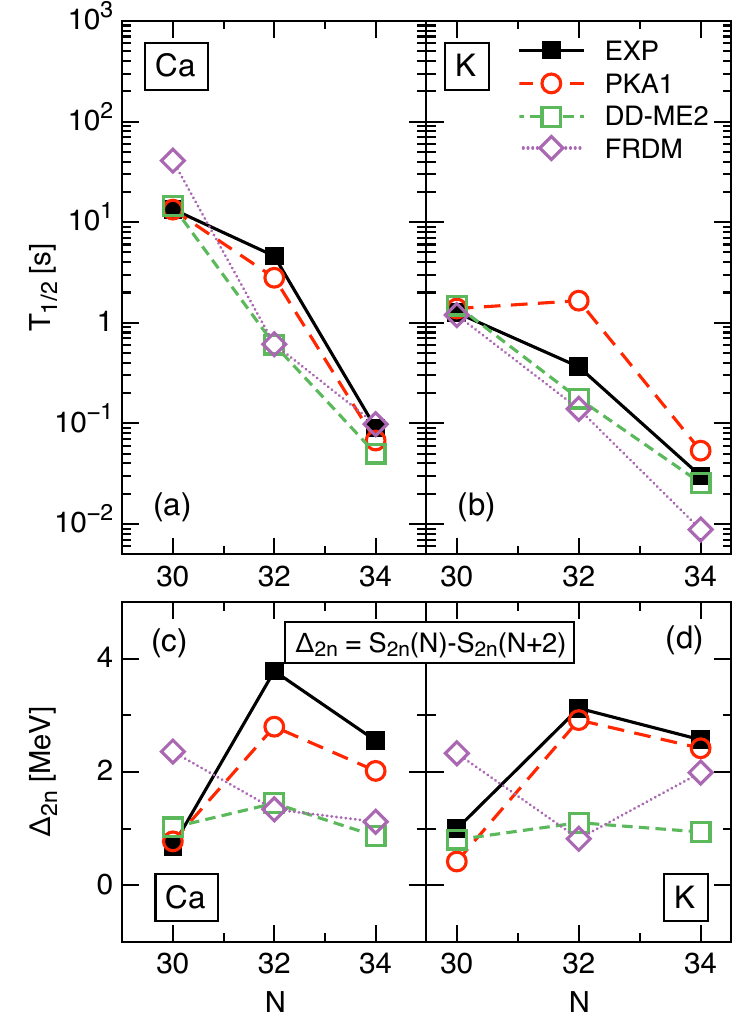}
	\vspace{-0.75cm}
	\begin{picture}(300,25)
	\end{picture}
	\caption{(a)(b) Calculated $\beta$-decay half-lives of Ca and K isotopes using the PKA1 and DD-ME2 \cite{Lalazissis2005} interactions. For the PKA1 interaction, the isoscalar pairing strength is fixed at $V_{0} = 200$ MeV for Ca isotopes and $V_{0} = 240$ MeV for K isotopes; for DD-ME2 interaction, it is $V_{0} = 80$ MeV for Ca and $V_{0} = 120$ MeV for K. The isovector pairing strength is set to its standard value $(f_{\text{IV}}= 1.0)$.  For comparison, experimental half‑lives \cite{Kondev2021} and FRDM \cite{Moeller2019} results are shown. (c)(d) The two-neutron gap, defined as the  difference of two-neutron separation energies, $\Delta_{2n}=S_{2n}(N)-S_{2n}(N+2)$, for  Ca and K isotopes calculated with the PKA1 and DD-ME2 interactions. Experimental values from the AME2020 evaluation \cite{Wang2021}, and FRDM predictions \cite{Moeller2019} are presented for comparison.}
	\label{fig1}
\end{figure}

First, we investigate the $\beta$-decay half-lives of Ca and K isotopes with $N = 30, 32$ and $34$. 
As shown in Fig. \ref{fig1} (a) and (b), the experimental half-lives show a gradual decrease from $N=30$ to $N=32$, followed by a more rapid decline from $N=32$ to $N=34$ for both Ca and K isotopes.  Notably, this reduction is more pronounced in Ca isotopes compared to K isotopes. 
The overall evolution trend from $N=30$ to $N=34$ is successfully reproduced by the PKA1 interaction; however, it overestimates the half-life of $^{51}$K ($N=32$), leading to %a worse reproduction of experimental trend in K isotopes compared to that in Ca isotopes.
 discrepancies in the detailed comparison with experimental data for Ca and K isotopes.
%In contrast, the DD-ME2 interaction shows a smooth decreasing trend, while the FRDM model exhibits an opposite trend compared to that observed in the experimental data.
In contrast, the DD-ME2 interaction and FRDM model exhibit a more rapid decrease at $N=32$ compared to that observed in the experimental data.

To assess the role of shell closure in shaping the half-life trend, we examine the two-neutron gap $\Delta_{2n}$ (defined as the difference in two-neutron separation energies) that reflects the shell gap magnitude.
Figure. \ref{fig1} (c) and (d) reveal that 
experimental results exhibit a significant increase in $\Delta_{2n}$ at $N = 32$ for Ca and K isotopes, which serves as the evidence for the existence of a shell closure.  The enhancement in Ca isotopes is stronger than in K isotopes. 
The PKA1 interaction successfully reproduces the enhancement at $N=32$, although it predicts a greater increase for K isotopes than for Ca isotopes.
%Conversely, the DD-ME2 interaction shows almost constant values of $\Delta_{2n}$ from $N=30$ to $N=34$, while the FRDM model predicts an opposite trend compared to the experiment.  
Conversely, the DD-ME2 interaction exhibits near-constant $\Delta_{2n}$ values from $N=30$ to $N=32$, whereas the FRDM model predicts a decrease  at $N = 32$.
These behaviors observed in shell-gap reproduction are consistent with those in $\beta$-decay half-lives. 
Comparing the variations in $\beta$-decay half-lives and $\Delta_{2n}$ for Ca and K isotopes, the longer half-lives observed at $N = 32$ in Ca and K isotopes are likely associated with the emergence of a new shell closure at this neutron number.
\begin{figure}[!t]	
	\setlength{\abovecaptionskip}{0.cm}
	\setlength{\belowcaptionskip}{-0.cm}  	
	\centering\includegraphics[width=1\linewidth]{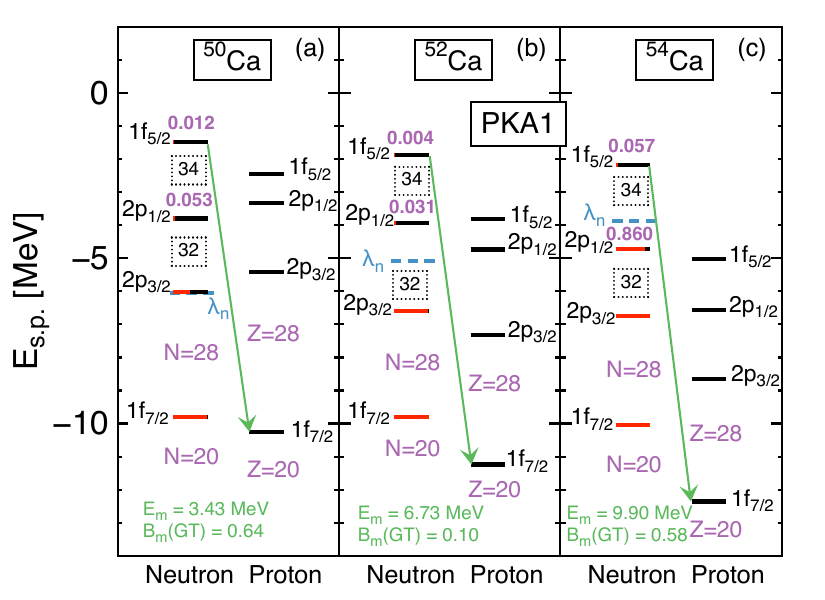}
	\vspace{-0.75cm}
	\begin{picture}(300,25)
	\end{picture}
	\caption{(a)(b)(c) The single-particle energy of $^{50,52,54}$Ca calculated using PKA1 interaction. Dominant GT transitions, critical for $\beta$-decay half-lives, are shown with their strengths [$B_m$(GT)] and $\beta$-decay transition energies ($E_m$) [cf. Eq. \eqref{Eq3}].
		The green arrow indicates the dominant single-particle GT transition in the $\beta$-decay half-life, namely $\nu 1f_{5/2} \rightarrow \pi 1f_{7/2}$. Additionally, the occupation probabilities of the $\nu 2p_{1/2}$ and $\nu 1f_{5/2}$ orbitals are displayed.  For more details, see the text.}
	\label{fig2}
\end{figure}

In order to elucidate the effect of shell structure on $\beta$-decay half-lives,  the detailed analysis on GT transitions that contribute most to $\beta$-decay is given based on the single-particle energy levels of $^{50,52,54}$Ca,  as shown in Fig. \ref{fig2}.
The dominant GT transitions for $\beta$-decay half-lives are presented in the figure by giving the specific values of transition strength $B_{m}$ and $\beta$-decay transition energy $E_{m}$ in Eq. \eqref{Eq3}. 
The dominant single-particle component in these GT transitions is $\nu 1f_{5/2} \rightarrow \pi 1f_{7/2}$ for these three nuclei, indicated by the green arrow. 
Moreover, the occupation probabilities of $\nu 2p_{1/2}$ and $\nu1f_{5/2}$ orbitals are also shown.
From $^{50}$Ca to $^{52}$Ca, the enhanced shell gap at $N = 32$ reduces the occupation probability of the $\nu 2p_{1/2}$ orbital, which in turn decreases the occupation probability in $\nu 1f_{5/2}$. The reduced occupation probability in  $\nu 1f_{5/2}$ orbital decreases the transition strength (from $B_{m}(\text{GT}) = 0.64$ to $0.10$). 
As a result, the decrease in half-lives, driven by the reduction in $\beta$-decay transition energy (from $ E_{m} = 3.43$ MeV to $6.73$ MeV), is seen to slow down. 
From $^{52}$Ca to $^{54}$Ca, despite the increased shell gap at $N = 34$, the occupation probability in the $\nu 1f_{5/2}$ rises as the $\nu 2p_{1/2}$ orbital approaches full occupation.
This increased occupancy leads to a more rapid decline in half-lives driven by the enhanced transition strength (from $B_{m}(\text{GT}) = 0.10$ to $0.58$) alongside the reduction in $\beta$-decay transition energy (from $ E_{m} = 6.73$ MeV to $9.90$ MeV).
Therefore, the occupation probability of $\nu 1f_{5/2}$ orbital determines the GT transition strength dominant in $\beta$-decay, which is 
 highly sensitive to the shell gap at $N =32 $, hence establishing a direct correlation between the $\beta$-decay half-life and shell gap.  Similar analysis can be applied to K isotopes as well, enabling analogous conclusions.

\begin{figure}[!t]	
	\setlength{\abovecaptionskip}{0.cm}
	\setlength{\belowcaptionskip}{-0.cm}  	
	\centering\includegraphics[width=1\linewidth]{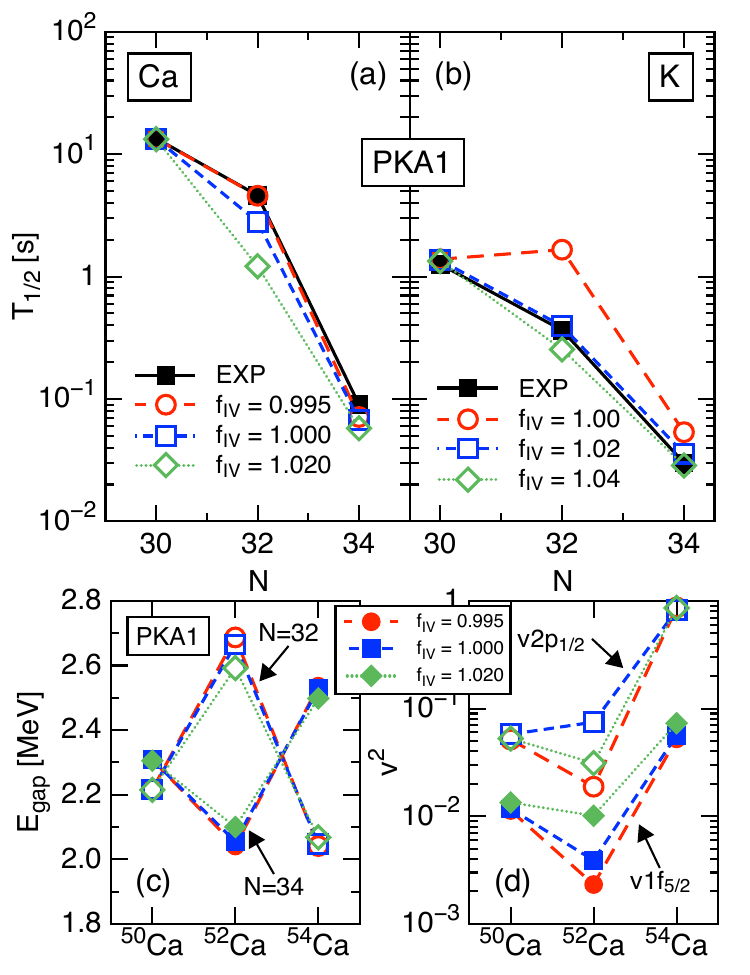}
	\vspace{-0.75cm}
	\begin{picture}(300,25)
	\end{picture}
	\caption{(a)(b) The $\beta$-decay half-lives of Ca and K isotopes calculated using PKA1 interaction with the different strength of isovector pairing force. The isoscalar pairing force strength is fixed at $V_{0} = 200$ MeV for Ca isotopes and $V_{0} = 240$ MeV for K isotopes. Experimental data \cite{Kondev2021} are included for comparison.  (c) The shell gaps at $N = 32$ and $N = 34$ for Ca isotopes,  defined as the difference between the corresponding single-particle energies,  are presented for different strength of isovector pairing force using the PKA1 interaction.  (d) The occupation probabilities for the $\nu 1f_{5/2}$ and $\nu 2p_{1/2}$ orbitals in Ca isotopes obtained by using the PKA1 interaction. }
	\label{fig3}
\end{figure}

A direct correspondence exists between occupation probability and shell gap in single-particle energy levels: a large shell gap suppress the occupation probability of states above the gap due to weak pairing scattering, whereas a small shell gap produces a high occupation probability in that level for the same reason.  
%The effect on the single-particle occupation probability by the shell gap can be mimic by the change of pairing strength.  The larger pairing strength makes the pairing scattering to the single-particle level above the shell gap easier, leading to a larger occupation probability, just the same effect as reducing the shell gap. 
This shell-gap effect on occupation probability can be equivalently described through pairing strength modulation. Increasing the pairing strength facilitates nucleons scattering into states above the shell gap, thereby elevating their occupation probabilities, an effect quantitatively similar to  reducing the shell gap itself.
Therefore, to further illustrate the connection between $\beta$-decay half-lives and shell gaps via occupation probabilities, one can vary the strengths of the isovector pairing force to mimic the change of shell gap, and investigate how these changes in occupation probabilities affect $\beta$-decay half-lives.
In Fig. \ref{fig3} (a) and (b), the $\beta$-decay half-lives calculated using PKA1 interaction with different strength of isovector pairing are presented.
For Ca and K isotopes, as the strength of isovector paring increases, the half-lives of nuclei with $N = 32$ ($^{52}$Ca and $^{51}$K) show noticeable decreases compared to their neighboring nuclei, which exhibit relatively minor changes. This behavior confirms the sensitivity of $\beta$-decay half-lives of $N = 32$ nuclei to $N=32$ shell gap. 
It is seen from  Fig. \ref{fig3}(c) and (d) that the  $\nu 1f_{5/2}$ orbital occupation probability in  $^{52}$Ca  is considerably lower than in adjacent nuclei due to the enhanced shell gap at $N=32$, and these reduced occupancies are particularly sensitive to variations in the isovector pairing strength or, equivalently, the shell gap magnitude, ultimately manifesting as noticeable differences in their half-lives.
%The enhanced $N = 32$ shell gap leads to lower occupation probability of $\nu 1f_{5/2}$ orbital in $^{52}$Ca compared to those in neighboring nuclei [see Fig. \ref{fig3} (c) and (d)], 
%which exhibit heightened sensitivity to the isovector pairing force, or equivalently to the shell gap,  leading to obvious variations in their half-lives.
%This behavior is attributed to the lower occupation probability of $\nu 1f_{5/2}$ orbital in $^{52}$Ca and $^{51}$K compared to those in neighboring nuclei, driven by the enhanced shell gap at $N = 32$ [see Fig. \ref{fig3} (c) and (d)].
%These low occupation probabilities exhibit heightened sensitivity to the isovector pairing force, or equivalently to the shell gap,  leading to obvious variations in their half-lives.
When the isovector pairing strength is sufficiently large, the occupation probability of the $\nu 1f_{5/2}$  
orbital shows no significant differences between $^{52}$Ca and $^{50}$Ca [see Fig. \ref{fig3}(d)]. Consequently, the half-lives of Ca isotopes exhibit an almost linear decreasing trend. This reproduces the scenario where $N=32$ is not a magic number, consistent with the DD-ME2 predictions.
The above phenomena can be also observed in K isotopes for the same reason.
%Therefore, the direct correlation between the half-lives and occupation probabilities is confirmed, with the magnitude of occupation probabilities reflecting the size of the shell gap.

From Fig. \ref{fig3} (a), it is evident that for Ca isotopes, the PKA1 interaction accurately reproduces experimental data when the strength of isovector pairing force is set to be 0.995.
This finding indicates that reproducing the data requires weaker isovector pairing, equivalent to a larger shell gap, than implied by the standard PKA1 setup. Consequently, it suggests that the shell gap predicted by PKA1 at 
$N=32$ is slightly smaller than observed experimentally. 
 %It means that weaker pairing scattering, or equivalently larger shell gap,  is needed to reproduce experimental data. 
%This suggests that the shell gap at $N = 32$ predicted by PKA1 is slightly smaller than experimental values. 
This is consistent with the results of $\Delta_{2n}$, where the theoretical change in $\Delta_{2n}$ is smaller than that observed experimentally from $N = 30$ to $32$ [see Fig. \ref{fig1} (c)].
In contrast, for K isotopes, PKA1 overestimates the half-life when using the standard strength of 1.00 for isovector pairing force.
However, it accurately  reproduces experimental data when the strength of isovector pairing force is increased to 1.02. 
This indicates the shell gap at $N = 32$ predicted by PKA1 is larger than experimental values. 
This is consistent with the results of $\Delta_{2n}$, where the theoretical change in $\Delta_{2n}$ is greater than that observed experimentally from $N = 30$ to $32$ [see Fig. \ref{fig1} (d)].
Moreover, the  trend of the decline in half-lives of K isotopes is closer to linear than that of Ca isotopes, indicating the corresponding shell gap at $N = 32$ is smaller in K isotopes compared with Ca isotopes.  The above findings of a strong $N=32$ shell gap in Ca isotopes and a weaker yet apparent gap in K isotopes are consistent with the experimental findings from mass \cite{ Rosenbusch2015,Wienholtz2013} and electromagnetic transition measurements \cite{Huck1985, Gade2006}. 
\begin{figure}[!t]	
	\setlength{\abovecaptionskip}{0.cm}
	\setlength{\belowcaptionskip}{-0.cm}  	
	\centering\includegraphics[width=1\linewidth]{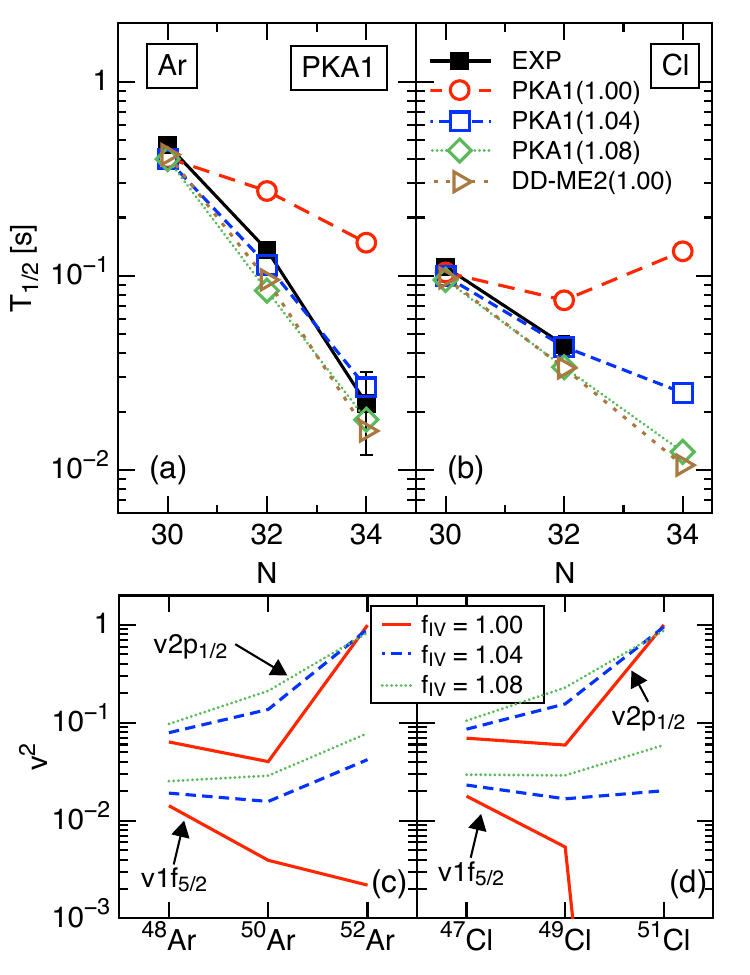}
	\vspace{-0.75cm}
	\begin{picture}(300,25)
	\end{picture}
	\caption{(a)(b) $\beta$-decay half-lives of Ar and Cl isotopes calculated using the PKA1 and DD-ME2 interactions with different isovector pairing strengths (values indicated in brackets). For the PKA1 interaction, the isoscalar pairing strength is fixed at $V_{0} = 255$ MeV for Ar isotopes and $V_{0} = 270$ MeV for Cl isotopes; for DD-ME2 interaction, it is $V_{0} = 175$ MeV for Ar and $V_{0} = 220$ MeV for Cl. Experimental data \cite{QBZeng2025} are provided for comparison.  (c)(d) Occupation probabilities for the $\nu 1f_{5/2}$ and $\nu 2p_{1/2}$ orbitals in Ar and Cl isotopes obtained with the PKA1 interaction using different isovector pairing strengths.  }
	\label{fig4}
\end{figure}

In order to investigate the shell structure at $N =32$  of nuclei with $Z \le 18$, the $\beta$-decay half-lives of Ar ($Z = 18$) and Cl ($Z = 17$) isotopes are shown in Fig. \ref{fig4} (a) and (b) calculated by PKA1 and DD-ME2 interactions with different isovector pairing strengths.
For the $\beta$-decay of Ar and Cl isotopes, transitions $\nu 1f_{5/2} \rightarrow \pi 1f_{7/2}$ and $\nu 2p_{1/2} \rightarrow \pi 2p_{3/2}$ predominantly characterize the decays of nuclei with $N = 32$ ($^{50}$Ar and $^{49}$Cl), while playing  important roles in the decays of nuclei with $N = 30$ and $34$ ($^{48,52}$Ar and $^{47,51}$Cl). 
For $N =32$ and $34$, as isovector pairing strength increases from $1.00$ to $1.04$, half-lives of Ar and Cl isotopes calculated using PKA1 interaction decrease significantly. 
This reduction is attributed to the markedly increased occupation probabilities in $\nu 1f_{5/2}$ and $\nu 2p_{1/2}$ orbitals [Fig. \ref{fig4} (c) and (d)], which are important for the corresponding GT transition strengths.
Note that the half-lives of Cl isotopes are increased from $N = 32$ to $34$ in PKA1 using the standard strength of 1.00 for isovector pairing force. 
This phenomenon occurs because PKA1 predicts a large shell gap at $N =34$, resulting in occupation probability in $\nu 1f_{5/2}$ orbital being close to zero in $^{51}$Cl. 
Consequently, the contribution from transition $\nu 1f_{5/2} \rightarrow \pi 1f_{7/2}$ vanishes in $^{51}$Cl, leading to a longer half-life at $N = 34$.

For Ar isotopes, PKA1 interaction successfully reproduces experimental half-lives when employing a pairing strengths of 1.04. 
As shown in Fig. \ref{fig4} (c), the occupation probabilities in $\nu 1f_{5/2}$ and $\nu 2p_{1/2}$ orbitals, at this pairing strength, are comparable for both $N = 30$ and $32$, suggesting that the shell gap at $N = 32$ is overestimated in PKA1 compared to experimental data. 
Furthermore, the results of PKA1 interaction with pairing strength of $1.04$ are consistent with that of DD-ME2 interaction, which does not predict the significant subshell at $N = 32$ in Ar isotopes. 
This indicates that a pronounced shell gap does not exist at $N = 32$ in Ar isotopes. 
In the case of Cl isotopes, using a pairing strengths of 1.04,  PKA1 interaction can reproduce experimental half-lives (with $N = 30$ and 32).
When the pairing strength is increased to 1.08, the half-life of $N=32$ doesn't change much, but the half-life of $N=34$ is further reduced due to an increased occupation in the $\nu 1f_{5/2}$ orbital, leading to similar results as DD-ME2 interaction.  
Although there is still no experimental data for $N=34$, the agreement with  $N=32$ data shows similar occupation probabilities in the $\nu 1f_{5/2}$ and $\nu 2p_{1/2}$ orbitals at $N=32$ compared to $N=30$, suggesting that no significant shell gap exists at $N = 32$ in Cl isotopes.
However, a robust confirmation necessitates further experimental measurements for Cl isotopes at $N=34$. 
%This conclusion is consistent with the recent experiment result \cite{QBZeng2025}.
%It is noteworthy that for the nucleus $^{51}$Cl (with $N = 34$), the half-life exhibits a change when the pairing strength is adjusted from 1.04 to 1.08, a variation attributed to an increased occupation in the $\nu 1f_{5/2}$ orbital [see Fig. \ref{fig4} (d)].

In the context of $\beta$-decay half-lives of nuclei near shell closures, the role of FF transitions may be important \cite{Borzov2003,Borzov2005,Yoshida2018,Yoshida2019}. 
In our calculations, we specifically evaluated this contribution by including FF transitions.
For Ca isotopes, the fully occupied $\pi 1d_{3/2}$ orbital eliminates possible FF transitions contributing to $\beta$-decay half-lives. 
In contrast, for K, Ar, and Cl isotopes, the partial occupation of the $\pi 1d_{3/2}$ orbital allows for possible FF transitions, including transitions $\nu 1f_{5/2}\rightarrow \pi 1d_{3/2}, \nu 2p_{1/2}\rightarrow \pi 1d_{3/2}, \nu 2p_{3/2}\rightarrow \pi 1d_{3/2}$, and $\nu 1f_{7/2}\rightarrow \pi 1d_{3/2}$. 
Despite this potential, it is found that these FF transitions remain subdominant relative to GT contributions, resulting in only minor changes to the calculated half-lives. Thus, the conclusions drawn from the allowed GT approximation continue to hold.

In summary, this work demonstrates that $\beta$-decay half-lives provide a novel probe of the magic number $N=32$. 
We establish a characteristic signature for a large shell gap through a gradual half-life decrease from $N = 30$ to 32, followed by a sharp decline from $N = 32$ to $N = 34$.
Self-consistent pnQRPA calculations reveal that the correlation between $\beta$-decay half-lives and the 
$N=32$ shell structure is mediated by occupation probabilities in orbitals above the shell gap, where the magnitude of orbital occupancy directly reflects the shell gap size. 
Crucially, this represents a completely novel mechanism: Here the reduced shell gaps permit pairing scattering to populate orbitals above the gap,  a behavior impossible in traditional magic number cases.
Our analysis confirms a robust $N=32$ shell gap in Ca isotopes and a weaker yet apparent gap in K isotopes, consistent with experimental mass and electromagnetic transition data. Conversely, no significant closed-shell evidence emerges at $N=32$ for Ar and Cl isotopes. This approach establishes a refined framework for investigating emergent magic numbers in exotic nuclei through nuclear weak-interaction processes.

%In summary, we have shown that the $\beta$-decay half-lives can serve as a new evidence for the new magic number $N=32$. It is found that the rapid decline from $N=32$ to $N=34$ after the gradual decrease from $N=30$ to $N=32$ is the signature for a big shell gap at $N=32$.  With the self-consistent pnQRPA theory, we illustrate that the correlation between $\beta$-decay half-lives and shell structure at $N = 32$ is linked through  the occupation probabilities in orbitals above the $N=32$ shell, where the magnitude of occupation probabilities reflecting the size of the shell gap.  Through the above analysis, we find that the shell gap at $N =32$ is strong in Ca isotopes, but diminishes in K isotopes, which is consistent with experiments through mass measurement and electromagnetic transitions. We show that the closed-shell structure at $ N = 32$  is not significant in Ar and Cl isotopes according to the present $\beta$-decay data. Our approach offers a refined perspective for investigating the new magic number structure through the weak interactions of atomic nuclei. 

%Our approach offers a refined perspective for investigating the new magic number structure at $N = 32$ through the weak interactions of atomic nuclei, and provides avenues for future research on the new magic number structure at $N = 34$.
%This insight not only deepens our understanding of nuclear shell structure and weak decay  processes, but also provides avenues for future research on the new magic number structure at $N = 34$.

The work is supported by the National Key Research and Development (R$\&$D) Program under Grant No. 2021YFA1601500, the National Natural Science Foundation of China (Grants No. 12075104, No.12447106, No. 12135004), the Lingchuang Research Project of China National Nuclear Corporation under Grant No.CNNC-LCKY-2024-082, 
and the Fundamental Research Funds for the Central Universities (Grants No. lzujbky-2024-it02, lzujbky-2023-stlt01).

\bibliography{ref}

%apsrev4-2.bst 2019-01-14 (MD) hand-edited version of apsrev4-1.bst
%Control: key (0)
%Control: author (72) initials jnrlst
%Control: editor formatted (1) identically to author
%Control: production of article title (-1) disabled
%Control: page (0) single
%Control: year (1) truncated
%Control: production of eprint (0) enabled
\begin{thebibliography}{72}%
\makeatletter
\providecommand \@ifxundefined [1]{%
 \@ifx{#1\undefined}
}%
\providecommand \@ifnum [1]{%
 \ifnum #1\expandafter \@firstoftwo
 \else \expandafter \@secondoftwo
 \fi
}%
\providecommand \@ifx [1]{%
 \ifx #1\expandafter \@firstoftwo
 \else \expandafter \@secondoftwo
 \fi
}%
\providecommand \natexlab [1]{#1}%
\providecommand \enquote  [1]{``#1''}%
\providecommand \bibnamefont  [1]{#1}%
\providecommand \bibfnamefont [1]{#1}%
\providecommand \citenamefont [1]{#1}%
\providecommand \href@noop [0]{\@secondoftwo}%
\providecommand \href [0]{\begingroup \@sanitize@url \@href}%
\providecommand \@href[1]{\@@startlink{#1}\@@href}%
\providecommand \@@href[1]{\endgroup#1\@@endlink}%
\providecommand \@sanitize@url [0]{\catcode `\\12\catcode `\$12\catcode
  `\&12\catcode `\#12\catcode `\^12\catcode `\_12\catcode `\%12\relax}%
\providecommand \@@startlink[1]{}%
\providecommand \@@endlink[0]{}%
\providecommand \url  [0]{\begingroup\@sanitize@url \@url }%
\providecommand \@url [1]{\endgroup\@href {#1}{\urlprefix }}%
\providecommand \urlprefix  [0]{URL }%
\providecommand \Eprint [0]{\href }%
\providecommand \doibase [0]{https://doi.org/}%
\providecommand \selectlanguage [0]{\@gobble}%
\providecommand \bibinfo  [0]{\@secondoftwo}%
\providecommand \bibfield  [0]{\@secondoftwo}%
\providecommand \translation [1]{[#1]}%
\providecommand \BibitemOpen [0]{}%
\providecommand \bibitemStop [0]{}%
\providecommand \bibitemNoStop [0]{.\EOS\space}%
\providecommand \EOS [0]{\spacefactor3000\relax}%
\providecommand \BibitemShut  [1]{\csname bibitem#1\endcsname}%
\let\auto@bib@innerbib\@empty
%</preamble>
\bibitem [{\citenamefont {Mayer}(1948)}]{Mayer1948}%
  \BibitemOpen
  \bibfield  {author} {\bibinfo {author} {\bibfnamefont {M.~G.}\ \bibnamefont
  {Mayer}},\ }\href {https://doi.org/10.1103/PhysRev.74.235} {\bibfield
  {journal} {\bibinfo  {journal} {Phys. Rev.}\ }\textbf {\bibinfo {volume}
  {74}},\ \bibinfo {pages} {235} (\bibinfo {year} {1948})}\BibitemShut
  {NoStop}%
\bibitem [{\citenamefont {Haxel}\ \emph {et~al.}(1949)\citenamefont {Haxel},
  \citenamefont {Jensen},\ and\ \citenamefont {Suess}}]{Haxel1949}%
  \BibitemOpen
  \bibfield  {author} {\bibinfo {author} {\bibfnamefont {O.}~\bibnamefont
  {Haxel}}, \bibinfo {author} {\bibfnamefont {J.~H.~D.}\ \bibnamefont
  {Jensen}},\ and\ \bibinfo {author} {\bibfnamefont {H.~E.}\ \bibnamefont
  {Suess}},\ }\href {https://doi.org/10.1103/PhysRev.75.1766.2} {\bibfield
  {journal} {\bibinfo  {journal} {Phys. Rev.}\ }\textbf {\bibinfo {volume}
  {75}},\ \bibinfo {pages} {1766} (\bibinfo {year} {1949})}\BibitemShut
  {NoStop}%
\bibitem [{\citenamefont {Otsuka}\ \emph {et~al.}(2001)\citenamefont {Otsuka},
  \citenamefont {Fujimoto}, \citenamefont {Utsuno}, \citenamefont {Brown},
  \citenamefont {Honma},\ and\ \citenamefont {Mizusaki}}]{Otsuka2001}%
  \BibitemOpen
  \bibfield  {author} {\bibinfo {author} {\bibfnamefont {T.}~\bibnamefont
  {Otsuka}}, \bibinfo {author} {\bibfnamefont {R.}~\bibnamefont {Fujimoto}},
  \bibinfo {author} {\bibfnamefont {Y.}~\bibnamefont {Utsuno}}, \bibinfo
  {author} {\bibfnamefont {B.~A.}\ \bibnamefont {Brown}}, \bibinfo {author}
  {\bibfnamefont {M.}~\bibnamefont {Honma}},\ and\ \bibinfo {author}
  {\bibfnamefont {T.}~\bibnamefont {Mizusaki}},\ }\href
  {https://doi.org/10.1103/PhysRevLett.87.082502} {\bibfield  {journal}
  {\bibinfo  {journal} {Phys. Rev. Lett.}\ }\textbf {\bibinfo {volume} {87}},\
  \bibinfo {pages} {082502} (\bibinfo {year} {2001})}\BibitemShut {NoStop}%
\bibitem [{\citenamefont {Meng}\ \emph {et~al.}(2006)\citenamefont {Meng},
  \citenamefont {Toki}, \citenamefont {Zhou}, \citenamefont {Zhang},
  \citenamefont {Long},\ and\ \citenamefont {Geng}}]{Meng2006}%
  \BibitemOpen
  \bibfield  {author} {\bibinfo {author} {\bibfnamefont {J.}~\bibnamefont
  {Meng}}, \bibinfo {author} {\bibfnamefont {H.}~\bibnamefont {Toki}}, \bibinfo
  {author} {\bibfnamefont {S.}~\bibnamefont {Zhou}}, \bibinfo {author}
  {\bibfnamefont {S.}~\bibnamefont {Zhang}}, \bibinfo {author} {\bibfnamefont
  {W.}~\bibnamefont {Long}},\ and\ \bibinfo {author} {\bibfnamefont
  {L.}~\bibnamefont {Geng}},\ }\href
  {https://doi.org/https://doi.org/10.1016/j.ppnp.2005.06.001} {\bibfield
  {journal} {\bibinfo  {journal} {Progress in Particle and Nuclear Physics}\
  }\textbf {\bibinfo {volume} {57}},\ \bibinfo {pages} {470} (\bibinfo {year}
  {2006})}\BibitemShut {NoStop}%
\bibitem [{\citenamefont {Sorlin}\ and\ \citenamefont
  {Porquet}(2008)}]{Sorlin2008}%
  \BibitemOpen
  \bibfield  {author} {\bibinfo {author} {\bibfnamefont {O.}~\bibnamefont
  {Sorlin}}\ and\ \bibinfo {author} {\bibfnamefont {M.-G.}\ \bibnamefont
  {Porquet}},\ }\href
  {https://doi.org/https://doi.org/10.1016/j.ppnp.2008.05.001} {\bibfield
  {journal} {\bibinfo  {journal} {Progress in Particle and Nuclear Physics}\
  }\textbf {\bibinfo {volume} {61}},\ \bibinfo {pages} {602} (\bibinfo {year}
  {2008})}\BibitemShut {NoStop}%
\bibitem [{\citenamefont {Simon}\ \emph {et~al.}(1999)\citenamefont {Simon},
  \citenamefont {Aleksandrov}, \citenamefont {Aumann}, \citenamefont
  {Axelsson}, \citenamefont {Baumann}, \citenamefont {Borge}, \citenamefont
  {Chulkov}, \citenamefont {Collatz}, \citenamefont {Cub}, \citenamefont
  {Dostal} \emph {et~al.}}]{Simon1999}%
  \BibitemOpen
  \bibfield  {author} {\bibinfo {author} {\bibfnamefont {H.}~\bibnamefont
  {Simon}}, \bibinfo {author} {\bibfnamefont {D.}~\bibnamefont {Aleksandrov}},
  \bibinfo {author} {\bibfnamefont {T.}~\bibnamefont {Aumann}}, \bibinfo
  {author} {\bibfnamefont {L.}~\bibnamefont {Axelsson}}, \bibinfo {author}
  {\bibfnamefont {T.}~\bibnamefont {Baumann}}, \bibinfo {author} {\bibfnamefont
  {M.~J.~G.}\ \bibnamefont {Borge}}, \bibinfo {author} {\bibfnamefont {L.~V.}\
  \bibnamefont {Chulkov}}, \bibinfo {author} {\bibfnamefont {R.}~\bibnamefont
  {Collatz}}, \bibinfo {author} {\bibfnamefont {J.}~\bibnamefont {Cub}},
  \bibinfo {author} {\bibfnamefont {W.}~\bibnamefont {Dostal}}, \emph
  {et~al.},\ }\href {https://doi.org/10.1103/PhysRevLett.83.496} {\bibfield
  {journal} {\bibinfo  {journal} {Phys. Rev. Lett.}\ }\textbf {\bibinfo
  {volume} {83}},\ \bibinfo {pages} {496} (\bibinfo {year} {1999})}\BibitemShut
  {NoStop}%
\bibitem [{\citenamefont {Motobayashi}\ \emph {et~al.}(1995)\citenamefont
  {Motobayashi}, \citenamefont {Ikeda}, \citenamefont {Ieki}, \citenamefont
  {Inoue}, \citenamefont {Iwasa}, \citenamefont {Kikuchi}, \citenamefont
  {Kurokawa}, \citenamefont {Moriya}, \citenamefont {Ogawa}, \citenamefont
  {Murakami}, \citenamefont {Shimoura}, \citenamefont {Yanagisawa},
  \citenamefont {Nakamura}, \citenamefont {Watanabe}, \citenamefont {Ishihara},
  \citenamefont {Teranishi}, \citenamefont {Okuno},\ and\ \citenamefont
  {Casten}}]{Motobayashi1995}%
  \BibitemOpen
  \bibfield  {author} {\bibinfo {author} {\bibfnamefont {T.}~\bibnamefont
  {Motobayashi}}, \bibinfo {author} {\bibfnamefont {Y.}~\bibnamefont {Ikeda}},
  \bibinfo {author} {\bibfnamefont {K.}~\bibnamefont {Ieki}}, \bibinfo {author}
  {\bibfnamefont {M.}~\bibnamefont {Inoue}}, \bibinfo {author} {\bibfnamefont
  {N.}~\bibnamefont {Iwasa}}, \bibinfo {author} {\bibfnamefont
  {T.}~\bibnamefont {Kikuchi}}, \bibinfo {author} {\bibfnamefont
  {M.}~\bibnamefont {Kurokawa}}, \bibinfo {author} {\bibfnamefont
  {S.}~\bibnamefont {Moriya}}, \bibinfo {author} {\bibfnamefont
  {S.}~\bibnamefont {Ogawa}}, \bibinfo {author} {\bibfnamefont
  {H.}~\bibnamefont {Murakami}}, \bibinfo {author} {\bibfnamefont
  {S.}~\bibnamefont {Shimoura}}, \bibinfo {author} {\bibfnamefont
  {Y.}~\bibnamefont {Yanagisawa}}, \bibinfo {author} {\bibfnamefont
  {T.}~\bibnamefont {Nakamura}}, \bibinfo {author} {\bibfnamefont
  {Y.}~\bibnamefont {Watanabe}}, \bibinfo {author} {\bibfnamefont
  {M.}~\bibnamefont {Ishihara}}, \bibinfo {author} {\bibfnamefont
  {T.}~\bibnamefont {Teranishi}}, \bibinfo {author} {\bibfnamefont
  {H.}~\bibnamefont {Okuno}},\ and\ \bibinfo {author} {\bibfnamefont
  {R.}~\bibnamefont {Casten}},\ }\href
  {https://doi.org/https://doi.org/10.1016/0370-2693(95)00012-A} {\bibfield
  {journal} {\bibinfo  {journal} {Physics Letters B}\ }\textbf {\bibinfo
  {volume} {346}},\ \bibinfo {pages} {9} (\bibinfo {year} {1995})}\BibitemShut
  {NoStop}%
\bibitem [{\citenamefont {Bastin}\ \emph {et~al.}(2007)\citenamefont {Bastin},
  \citenamefont {Gr\'evy}, \citenamefont {Sohler}, \citenamefont {Sorlin},
  \citenamefont {Dombr\'adi}, \citenamefont {Achouri}, \citenamefont
  {Ang\'elique}, \citenamefont {Azaiez}, \citenamefont {Baiborodin},
  \citenamefont {Borcea} \emph {et~al.}}]{Bastin2007}%
  \BibitemOpen
  \bibfield  {author} {\bibinfo {author} {\bibfnamefont {B.}~\bibnamefont
  {Bastin}}, \bibinfo {author} {\bibfnamefont {S.}~\bibnamefont {Gr\'evy}},
  \bibinfo {author} {\bibfnamefont {D.}~\bibnamefont {Sohler}}, \bibinfo
  {author} {\bibfnamefont {O.}~\bibnamefont {Sorlin}}, \bibinfo {author}
  {\bibfnamefont {Z.}~\bibnamefont {Dombr\'adi}}, \bibinfo {author}
  {\bibfnamefont {N.~L.}\ \bibnamefont {Achouri}}, \bibinfo {author}
  {\bibfnamefont {J.~C.}\ \bibnamefont {Ang\'elique}}, \bibinfo {author}
  {\bibfnamefont {F.}~\bibnamefont {Azaiez}}, \bibinfo {author} {\bibfnamefont
  {D.}~\bibnamefont {Baiborodin}}, \bibinfo {author} {\bibfnamefont
  {R.}~\bibnamefont {Borcea}}, \emph {et~al.},\ }\href
  {https://doi.org/10.1103/PhysRevLett.99.022503} {\bibfield  {journal}
  {\bibinfo  {journal} {Phys. Rev. Lett.}\ }\textbf {\bibinfo {volume} {99}},\
  \bibinfo {pages} {022503} (\bibinfo {year} {2007})}\BibitemShut {NoStop}%
\bibitem [{\citenamefont {Ozawa}\ \emph {et~al.}(2000)\citenamefont {Ozawa},
  \citenamefont {Kobayashi}, \citenamefont {Suzuki}, \citenamefont {Yoshida},\
  and\ \citenamefont {Tanihata}}]{Ozawa2000}%
  \BibitemOpen
  \bibfield  {author} {\bibinfo {author} {\bibfnamefont {A.}~\bibnamefont
  {Ozawa}}, \bibinfo {author} {\bibfnamefont {T.}~\bibnamefont {Kobayashi}},
  \bibinfo {author} {\bibfnamefont {T.}~\bibnamefont {Suzuki}}, \bibinfo
  {author} {\bibfnamefont {K.}~\bibnamefont {Yoshida}},\ and\ \bibinfo {author}
  {\bibfnamefont {I.}~\bibnamefont {Tanihata}},\ }\href
  {https://doi.org/10.1103/PhysRevLett.84.5493} {\bibfield  {journal} {\bibinfo
   {journal} {Phys. Rev. Lett.}\ }\textbf {\bibinfo {volume} {84}},\ \bibinfo
  {pages} {5493} (\bibinfo {year} {2000})}\BibitemShut {NoStop}%
\bibitem [{\citenamefont {Stanoiu}\ \emph {et~al.}(2004)\citenamefont
  {Stanoiu}, \citenamefont {Azaiez}, \citenamefont {Dombr\'adi}, \citenamefont
  {Sorlin}, \citenamefont {Brown}, \citenamefont {Belleguic}, \citenamefont
  {Sohler}, \citenamefont {Saint~Laurent}, \citenamefont {Lopez-Jimenez},
  \citenamefont {Penionzhkevich} \emph {et~al.}}]{Stanoiu2004}%
  \BibitemOpen
  \bibfield  {author} {\bibinfo {author} {\bibfnamefont {M.}~\bibnamefont
  {Stanoiu}}, \bibinfo {author} {\bibfnamefont {F.}~\bibnamefont {Azaiez}},
  \bibinfo {author} {\bibfnamefont {Z.}~\bibnamefont {Dombr\'adi}}, \bibinfo
  {author} {\bibfnamefont {O.}~\bibnamefont {Sorlin}}, \bibinfo {author}
  {\bibfnamefont {B.~A.}\ \bibnamefont {Brown}}, \bibinfo {author}
  {\bibfnamefont {M.}~\bibnamefont {Belleguic}}, \bibinfo {author}
  {\bibfnamefont {D.}~\bibnamefont {Sohler}}, \bibinfo {author} {\bibfnamefont
  {M.~G.}\ \bibnamefont {Saint~Laurent}}, \bibinfo {author} {\bibfnamefont
  {M.~J.}\ \bibnamefont {Lopez-Jimenez}}, \bibinfo {author} {\bibfnamefont
  {Y.~E.}\ \bibnamefont {Penionzhkevich}}, \emph {et~al.},\ }\href
  {https://doi.org/10.1103/PhysRevC.69.034312} {\bibfield  {journal} {\bibinfo
  {journal} {Phys. Rev. C}\ }\textbf {\bibinfo {volume} {69}},\ \bibinfo
  {pages} {034312} (\bibinfo {year} {2004})}\BibitemShut {NoStop}%
\bibitem [{\citenamefont {Doornenbal}\ \emph {et~al.}(2007)\citenamefont
  {Doornenbal}, \citenamefont {Reiter}, \citenamefont {Grawe}, \citenamefont
  {Otsuka}, \citenamefont {Al-Khatib}, \citenamefont {Banu}, \citenamefont
  {Beck}, \citenamefont {Becker}, \citenamefont {Bednarczyk}, \citenamefont
  {Benzoni} \emph {et~al.}}]{Doornenbal2007}%
  \BibitemOpen
  \bibfield  {author} {\bibinfo {author} {\bibfnamefont {P.}~\bibnamefont
  {Doornenbal}}, \bibinfo {author} {\bibfnamefont {P.}~\bibnamefont {Reiter}},
  \bibinfo {author} {\bibfnamefont {H.}~\bibnamefont {Grawe}}, \bibinfo
  {author} {\bibfnamefont {T.}~\bibnamefont {Otsuka}}, \bibinfo {author}
  {\bibfnamefont {A.}~\bibnamefont {Al-Khatib}}, \bibinfo {author}
  {\bibfnamefont {A.}~\bibnamefont {Banu}}, \bibinfo {author} {\bibfnamefont
  {T.}~\bibnamefont {Beck}}, \bibinfo {author} {\bibfnamefont {F.}~\bibnamefont
  {Becker}}, \bibinfo {author} {\bibfnamefont {P.}~\bibnamefont {Bednarczyk}},
  \bibinfo {author} {\bibfnamefont {G.}~\bibnamefont {Benzoni}}, \emph
  {et~al.},\ }\href
  {https://doi.org/https://doi.org/10.1016/j.physletb.2007.02.001} {\bibfield
  {journal} {\bibinfo  {journal} {Physics Letters B}\ }\textbf {\bibinfo
  {volume} {647}},\ \bibinfo {pages} {237} (\bibinfo {year}
  {2007})}\BibitemShut {NoStop}%
\bibitem [{\citenamefont {Hoffman}\ \emph {et~al.}(2008)\citenamefont
  {Hoffman}, \citenamefont {Baumann}, \citenamefont {Bazin}, \citenamefont
  {Brown}, \citenamefont {Christian}, \citenamefont {DeYoung}, \citenamefont
  {Finck}, \citenamefont {Frank}, \citenamefont {Hinnefeld}, \citenamefont
  {Howes} \emph {et~al.}}]{Hoffman2008}%
  \BibitemOpen
  \bibfield  {author} {\bibinfo {author} {\bibfnamefont {C.~R.}\ \bibnamefont
  {Hoffman}}, \bibinfo {author} {\bibfnamefont {T.}~\bibnamefont {Baumann}},
  \bibinfo {author} {\bibfnamefont {D.}~\bibnamefont {Bazin}}, \bibinfo
  {author} {\bibfnamefont {J.}~\bibnamefont {Brown}}, \bibinfo {author}
  {\bibfnamefont {G.}~\bibnamefont {Christian}}, \bibinfo {author}
  {\bibfnamefont {P.~A.}\ \bibnamefont {DeYoung}}, \bibinfo {author}
  {\bibfnamefont {J.~E.}\ \bibnamefont {Finck}}, \bibinfo {author}
  {\bibfnamefont {N.}~\bibnamefont {Frank}}, \bibinfo {author} {\bibfnamefont
  {J.}~\bibnamefont {Hinnefeld}}, \bibinfo {author} {\bibfnamefont
  {R.}~\bibnamefont {Howes}}, \emph {et~al.},\ }\href
  {https://doi.org/10.1103/PhysRevLett.100.152502} {\bibfield  {journal}
  {\bibinfo  {journal} {Phys. Rev. Lett.}\ }\textbf {\bibinfo {volume} {100}},\
  \bibinfo {pages} {152502} (\bibinfo {year} {2008})}\BibitemShut {NoStop}%
\bibitem [{\citenamefont {Kanungo}\ \emph {et~al.}(2009)\citenamefont
  {Kanungo}, \citenamefont {Nociforo}, \citenamefont {Prochazka}, \citenamefont
  {Aumann}, \citenamefont {Boutin}, \citenamefont {Cortina-Gil}, \citenamefont
  {Davids}, \citenamefont {Diakaki}, \citenamefont {Farinon}, \citenamefont
  {Geissel} \emph {et~al.}}]{Kanungo2009}%
  \BibitemOpen
  \bibfield  {author} {\bibinfo {author} {\bibfnamefont {R.}~\bibnamefont
  {Kanungo}}, \bibinfo {author} {\bibfnamefont {C.}~\bibnamefont {Nociforo}},
  \bibinfo {author} {\bibfnamefont {A.}~\bibnamefont {Prochazka}}, \bibinfo
  {author} {\bibfnamefont {T.}~\bibnamefont {Aumann}}, \bibinfo {author}
  {\bibfnamefont {D.}~\bibnamefont {Boutin}}, \bibinfo {author} {\bibfnamefont
  {D.}~\bibnamefont {Cortina-Gil}}, \bibinfo {author} {\bibfnamefont
  {B.}~\bibnamefont {Davids}}, \bibinfo {author} {\bibfnamefont
  {M.}~\bibnamefont {Diakaki}}, \bibinfo {author} {\bibfnamefont
  {F.}~\bibnamefont {Farinon}}, \bibinfo {author} {\bibfnamefont
  {H.}~\bibnamefont {Geissel}}, \emph {et~al.},\ }\href
  {https://doi.org/10.1103/PhysRevLett.102.152501} {\bibfield  {journal}
  {\bibinfo  {journal} {Phys. Rev. Lett.}\ }\textbf {\bibinfo {volume} {102}},\
  \bibinfo {pages} {152501} (\bibinfo {year} {2009})}\BibitemShut {NoStop}%
\bibitem [{\citenamefont {Tshoo}\ \emph {et~al.}(2012)\citenamefont {Tshoo},
  \citenamefont {Satou}, \citenamefont {Bhang}, \citenamefont {Choi},
  \citenamefont {Nakamura}, \citenamefont {Kondo}, \citenamefont {Deguchi},
  \citenamefont {Kawada}, \citenamefont {Kobayashi}, \citenamefont {Nakayama}
  \emph {et~al.}}]{Tshoo2012}%
  \BibitemOpen
  \bibfield  {author} {\bibinfo {author} {\bibfnamefont {K.}~\bibnamefont
  {Tshoo}}, \bibinfo {author} {\bibfnamefont {Y.}~\bibnamefont {Satou}},
  \bibinfo {author} {\bibfnamefont {H.}~\bibnamefont {Bhang}}, \bibinfo
  {author} {\bibfnamefont {S.}~\bibnamefont {Choi}}, \bibinfo {author}
  {\bibfnamefont {T.}~\bibnamefont {Nakamura}}, \bibinfo {author}
  {\bibfnamefont {Y.}~\bibnamefont {Kondo}}, \bibinfo {author} {\bibfnamefont
  {S.}~\bibnamefont {Deguchi}}, \bibinfo {author} {\bibfnamefont
  {Y.}~\bibnamefont {Kawada}}, \bibinfo {author} {\bibfnamefont
  {N.}~\bibnamefont {Kobayashi}}, \bibinfo {author} {\bibfnamefont
  {Y.}~\bibnamefont {Nakayama}}, \emph {et~al.},\ }\href
  {https://doi.org/10.1103/PhysRevLett.109.022501} {\bibfield  {journal}
  {\bibinfo  {journal} {Phys. Rev. Lett.}\ }\textbf {\bibinfo {volume} {109}},\
  \bibinfo {pages} {022501} (\bibinfo {year} {2012})}\BibitemShut {NoStop}%
\bibitem [{\citenamefont {Tanihata}\ \emph {et~al.}(2013)\citenamefont
  {Tanihata}, \citenamefont {Savajols},\ and\ \citenamefont
  {Kanungo}}]{Tanihata2013}%
  \BibitemOpen
  \bibfield  {author} {\bibinfo {author} {\bibfnamefont {I.}~\bibnamefont
  {Tanihata}}, \bibinfo {author} {\bibfnamefont {H.}~\bibnamefont {Savajols}},\
  and\ \bibinfo {author} {\bibfnamefont {R.}~\bibnamefont {Kanungo}},\ }\href
  {https://doi.org/https://doi.org/10.1016/j.ppnp.2012.07.001} {\bibfield
  {journal} {\bibinfo  {journal} {Progress in Particle and Nuclear Physics}\
  }\textbf {\bibinfo {volume} {68}},\ \bibinfo {pages} {215} (\bibinfo {year}
  {2013})}\BibitemShut {NoStop}%
\bibitem [{\citenamefont {Otsuka}\ \emph {et~al.}(2020)\citenamefont {Otsuka},
  \citenamefont {Gade}, \citenamefont {Sorlin}, \citenamefont {Suzuki},\ and\
  \citenamefont {Utsuno}}]{Otsuka2020}%
  \BibitemOpen
  \bibfield  {author} {\bibinfo {author} {\bibfnamefont {T.}~\bibnamefont
  {Otsuka}}, \bibinfo {author} {\bibfnamefont {A.}~\bibnamefont {Gade}},
  \bibinfo {author} {\bibfnamefont {O.}~\bibnamefont {Sorlin}}, \bibinfo
  {author} {\bibfnamefont {T.}~\bibnamefont {Suzuki}},\ and\ \bibinfo {author}
  {\bibfnamefont {Y.}~\bibnamefont {Utsuno}},\ }\href
  {https://doi.org/10.1103/RevModPhys.92.015002} {\bibfield  {journal}
  {\bibinfo  {journal} {Rev. Mod. Phys.}\ }\textbf {\bibinfo {volume} {92}},\
  \bibinfo {pages} {015002} (\bibinfo {year} {2020})}\BibitemShut {NoStop}%
\bibitem [{\citenamefont {Huck}\ \emph {et~al.}(1985)\citenamefont {Huck},
  \citenamefont {Klotz}, \citenamefont {Knipper}, \citenamefont {Mieh\'e},
  \citenamefont {Richard-Serre}, \citenamefont {Walter}, \citenamefont {Poves},
  \citenamefont {Ravn},\ and\ \citenamefont {Marguier}}]{Huck1985}%
  \BibitemOpen
  \bibfield  {author} {\bibinfo {author} {\bibfnamefont {A.}~\bibnamefont
  {Huck}}, \bibinfo {author} {\bibfnamefont {G.}~\bibnamefont {Klotz}},
  \bibinfo {author} {\bibfnamefont {A.}~\bibnamefont {Knipper}}, \bibinfo
  {author} {\bibfnamefont {C.}~\bibnamefont {Mieh\'e}}, \bibinfo {author}
  {\bibfnamefont {C.}~\bibnamefont {Richard-Serre}}, \bibinfo {author}
  {\bibfnamefont {G.}~\bibnamefont {Walter}}, \bibinfo {author} {\bibfnamefont
  {A.}~\bibnamefont {Poves}}, \bibinfo {author} {\bibfnamefont {H.~L.}\
  \bibnamefont {Ravn}},\ and\ \bibinfo {author} {\bibfnamefont
  {G.}~\bibnamefont {Marguier}},\ }\href
  {https://doi.org/10.1103/PhysRevC.31.2226} {\bibfield  {journal} {\bibinfo
  {journal} {Phys. Rev. C}\ }\textbf {\bibinfo {volume} {31}},\ \bibinfo
  {pages} {2226} (\bibinfo {year} {1985})}\BibitemShut {NoStop}%
\bibitem [{\citenamefont {Gade}\ \emph {et~al.}(2006)\citenamefont {Gade},
  \citenamefont {Janssens}, \citenamefont {Bazin}, \citenamefont {Broda},
  \citenamefont {Brown}, \citenamefont {Campbell}, \citenamefont {Carpenter},
  \citenamefont {Cook}, \citenamefont {Deacon}, \citenamefont {Dinca} \emph
  {et~al.}}]{Gade2006}%
  \BibitemOpen
  \bibfield  {author} {\bibinfo {author} {\bibfnamefont {A.}~\bibnamefont
  {Gade}}, \bibinfo {author} {\bibfnamefont {R.~V.~F.}\ \bibnamefont
  {Janssens}}, \bibinfo {author} {\bibfnamefont {D.}~\bibnamefont {Bazin}},
  \bibinfo {author} {\bibfnamefont {R.}~\bibnamefont {Broda}}, \bibinfo
  {author} {\bibfnamefont {B.~A.}\ \bibnamefont {Brown}}, \bibinfo {author}
  {\bibfnamefont {C.~M.}\ \bibnamefont {Campbell}}, \bibinfo {author}
  {\bibfnamefont {M.~P.}\ \bibnamefont {Carpenter}}, \bibinfo {author}
  {\bibfnamefont {J.~M.}\ \bibnamefont {Cook}}, \bibinfo {author}
  {\bibfnamefont {A.~N.}\ \bibnamefont {Deacon}}, \bibinfo {author}
  {\bibfnamefont {D.-C.}\ \bibnamefont {Dinca}}, \emph {et~al.},\ }\href
  {https://doi.org/10.1103/PhysRevC.74.021302} {\bibfield  {journal} {\bibinfo
  {journal} {Phys. Rev. C}\ }\textbf {\bibinfo {volume} {74}},\ \bibinfo
  {pages} {021302} (\bibinfo {year} {2006})}\BibitemShut {NoStop}%
\bibitem [{\citenamefont {Janssens}\ \emph {et~al.}(2002)\citenamefont
  {Janssens}, \citenamefont {Fornal}, \citenamefont {Mantica}, \citenamefont
  {Brown}, \citenamefont {Broda}, \citenamefont {Bhattacharyya}, \citenamefont
  {Carpenter}, \citenamefont {Cinausero}, \citenamefont {Daly}, \citenamefont
  {Davies} \emph {et~al.}}]{Janssens2002}%
  \BibitemOpen
  \bibfield  {author} {\bibinfo {author} {\bibfnamefont {R.}~\bibnamefont
  {Janssens}}, \bibinfo {author} {\bibfnamefont {B.}~\bibnamefont {Fornal}},
  \bibinfo {author} {\bibfnamefont {P.}~\bibnamefont {Mantica}}, \bibinfo
  {author} {\bibfnamefont {B.}~\bibnamefont {Brown}}, \bibinfo {author}
  {\bibfnamefont {R.}~\bibnamefont {Broda}}, \bibinfo {author} {\bibfnamefont
  {P.}~\bibnamefont {Bhattacharyya}}, \bibinfo {author} {\bibfnamefont
  {M.}~\bibnamefont {Carpenter}}, \bibinfo {author} {\bibfnamefont
  {M.}~\bibnamefont {Cinausero}}, \bibinfo {author} {\bibfnamefont
  {P.}~\bibnamefont {Daly}}, \bibinfo {author} {\bibfnamefont {A.}~\bibnamefont
  {Davies}}, \emph {et~al.},\ }\href
  {https://doi.org/https://doi.org/10.1016/S0370-2693(02)02682-5} {\bibfield
  {journal} {\bibinfo  {journal} {Physics Letters B}\ }\textbf {\bibinfo
  {volume} {546}},\ \bibinfo {pages} {55} (\bibinfo {year} {2002})}\BibitemShut
  {NoStop}%
\bibitem [{\citenamefont {Dinca}\ \emph {et~al.}(2005)\citenamefont {Dinca},
  \citenamefont {Janssens}, \citenamefont {Gade}, \citenamefont {Bazin},
  \citenamefont {Broda}, \citenamefont {Brown}, \citenamefont {Campbell},
  \citenamefont {Carpenter}, \citenamefont {Chowdhury}, \citenamefont {Cook}
  \emph {et~al.}}]{Dinca2005}%
  \BibitemOpen
  \bibfield  {author} {\bibinfo {author} {\bibfnamefont {D.-C.}\ \bibnamefont
  {Dinca}}, \bibinfo {author} {\bibfnamefont {R.~V.~F.}\ \bibnamefont
  {Janssens}}, \bibinfo {author} {\bibfnamefont {A.}~\bibnamefont {Gade}},
  \bibinfo {author} {\bibfnamefont {D.}~\bibnamefont {Bazin}}, \bibinfo
  {author} {\bibfnamefont {R.}~\bibnamefont {Broda}}, \bibinfo {author}
  {\bibfnamefont {B.~A.}\ \bibnamefont {Brown}}, \bibinfo {author}
  {\bibfnamefont {C.~M.}\ \bibnamefont {Campbell}}, \bibinfo {author}
  {\bibfnamefont {M.~P.}\ \bibnamefont {Carpenter}}, \bibinfo {author}
  {\bibfnamefont {P.}~\bibnamefont {Chowdhury}}, \bibinfo {author}
  {\bibfnamefont {J.~M.}\ \bibnamefont {Cook}}, \emph {et~al.},\ }\href
  {https://doi.org/10.1103/PhysRevC.71.041302} {\bibfield  {journal} {\bibinfo
  {journal} {Phys. Rev. C}\ }\textbf {\bibinfo {volume} {71}},\ \bibinfo
  {pages} {041302} (\bibinfo {year} {2005})}\BibitemShut {NoStop}%
\bibitem [{\citenamefont {Prisciandaro}\ \emph {et~al.}(2001)\citenamefont
  {Prisciandaro}, \citenamefont {Mantica}, \citenamefont {Brown}, \citenamefont
  {Anthony}, \citenamefont {Cooper}, \citenamefont {Garcia}, \citenamefont
  {Groh}, \citenamefont {Komives}, \citenamefont {Kumarasiri}, \citenamefont
  {Lofy}, \citenamefont {Oros-Peusquens}, \citenamefont {Tabor},\ and\
  \citenamefont {Wiedeking}}]{Prisciandaro2001}%
  \BibitemOpen
  \bibfield  {author} {\bibinfo {author} {\bibfnamefont {J.}~\bibnamefont
  {Prisciandaro}}, \bibinfo {author} {\bibfnamefont {P.}~\bibnamefont
  {Mantica}}, \bibinfo {author} {\bibfnamefont {B.}~\bibnamefont {Brown}},
  \bibinfo {author} {\bibfnamefont {D.}~\bibnamefont {Anthony}}, \bibinfo
  {author} {\bibfnamefont {M.}~\bibnamefont {Cooper}}, \bibinfo {author}
  {\bibfnamefont {A.}~\bibnamefont {Garcia}}, \bibinfo {author} {\bibfnamefont
  {D.}~\bibnamefont {Groh}}, \bibinfo {author} {\bibfnamefont {A.}~\bibnamefont
  {Komives}}, \bibinfo {author} {\bibfnamefont {W.}~\bibnamefont {Kumarasiri}},
  \bibinfo {author} {\bibfnamefont {P.}~\bibnamefont {Lofy}}, \bibinfo {author}
  {\bibfnamefont {A.}~\bibnamefont {Oros-Peusquens}}, \bibinfo {author}
  {\bibfnamefont {S.}~\bibnamefont {Tabor}},\ and\ \bibinfo {author}
  {\bibfnamefont {M.}~\bibnamefont {Wiedeking}},\ }\href
  {https://doi.org/https://doi.org/10.1016/S0370-2693(01)00565-2} {\bibfield
  {journal} {\bibinfo  {journal} {Physics Letters B}\ }\textbf {\bibinfo
  {volume} {510}},\ \bibinfo {pages} {17} (\bibinfo {year} {2001})}\BibitemShut
  {NoStop}%
\bibitem [{\citenamefont {B\"urger}\ \emph {et~al.}(2005)\citenamefont
  {B\"urger}, \citenamefont {Saito}, \citenamefont {Grawe}, \citenamefont
  {H\"ubel}, \citenamefont {Reiter}, \citenamefont {Gerl}, \citenamefont
  {G\'orska}, \citenamefont {Wollersheim}, \citenamefont {Al-Khatib},
  \citenamefont {Banu} \emph {et~al.}}]{Buerger2005}%
  \BibitemOpen
  \bibfield  {author} {\bibinfo {author} {\bibfnamefont {A.}~\bibnamefont
  {B\"urger}}, \bibinfo {author} {\bibfnamefont {T.}~\bibnamefont {Saito}},
  \bibinfo {author} {\bibfnamefont {H.}~\bibnamefont {Grawe}}, \bibinfo
  {author} {\bibfnamefont {H.}~\bibnamefont {H\"ubel}}, \bibinfo {author}
  {\bibfnamefont {P.}~\bibnamefont {Reiter}}, \bibinfo {author} {\bibfnamefont
  {J.}~\bibnamefont {Gerl}}, \bibinfo {author} {\bibfnamefont {M.}~\bibnamefont
  {G\'orska}}, \bibinfo {author} {\bibfnamefont {H.}~\bibnamefont
  {Wollersheim}}, \bibinfo {author} {\bibfnamefont {A.}~\bibnamefont
  {Al-Khatib}}, \bibinfo {author} {\bibfnamefont {A.}~\bibnamefont {Banu}},
  \emph {et~al.},\ }\href
  {https://doi.org/https://doi.org/10.1016/j.physletb.2005.07.004} {\bibfield
  {journal} {\bibinfo  {journal} {Physics Letters B}\ }\textbf {\bibinfo
  {volume} {622}},\ \bibinfo {pages} {29} (\bibinfo {year} {2005})}\BibitemShut
  {NoStop}%
\bibitem [{\citenamefont {Rosenbusch}\ \emph {et~al.}(2015)\citenamefont
  {Rosenbusch}, \citenamefont {Ascher}, \citenamefont {Atanasov}, \citenamefont
  {Barbieri}, \citenamefont {Beck}, \citenamefont {Blaum}, \citenamefont
  {Borgmann}, \citenamefont {Breitenfeldt}, \citenamefont {Cakirli},
  \citenamefont {Cipollone} \emph {et~al.}}]{Rosenbusch2015}%
  \BibitemOpen
  \bibfield  {author} {\bibinfo {author} {\bibfnamefont {M.}~\bibnamefont
  {Rosenbusch}}, \bibinfo {author} {\bibfnamefont {P.}~\bibnamefont {Ascher}},
  \bibinfo {author} {\bibfnamefont {D.}~\bibnamefont {Atanasov}}, \bibinfo
  {author} {\bibfnamefont {C.}~\bibnamefont {Barbieri}}, \bibinfo {author}
  {\bibfnamefont {D.}~\bibnamefont {Beck}}, \bibinfo {author} {\bibfnamefont
  {K.}~\bibnamefont {Blaum}}, \bibinfo {author} {\bibfnamefont
  {C.}~\bibnamefont {Borgmann}}, \bibinfo {author} {\bibfnamefont
  {M.}~\bibnamefont {Breitenfeldt}}, \bibinfo {author} {\bibfnamefont {R.~B.}\
  \bibnamefont {Cakirli}}, \bibinfo {author} {\bibfnamefont {A.}~\bibnamefont
  {Cipollone}}, \emph {et~al.},\ }\href
  {https://doi.org/10.1103/PhysRevLett.114.202501} {\bibfield  {journal}
  {\bibinfo  {journal} {Phys. Rev. Lett.}\ }\textbf {\bibinfo {volume} {114}},\
  \bibinfo {pages} {202501} (\bibinfo {year} {2015})}\BibitemShut {NoStop}%
\bibitem [{\citenamefont {Wienholtz}\ \emph {et~al.}(2013)\citenamefont
  {Wienholtz}, \citenamefont {Beck}, \citenamefont {Blaum}, \citenamefont
  {Borgmann}, \citenamefont {Breitenfeldt}, \citenamefont {Cakirli},
  \citenamefont {George}, \citenamefont {Herfurth}, \citenamefont {Holt},
  \citenamefont {Kowalska} \emph {et~al.}}]{Wienholtz2013}%
  \BibitemOpen
  \bibfield  {author} {\bibinfo {author} {\bibfnamefont {F.}~\bibnamefont
  {Wienholtz}}, \bibinfo {author} {\bibfnamefont {D.}~\bibnamefont {Beck}},
  \bibinfo {author} {\bibfnamefont {K.}~\bibnamefont {Blaum}}, \bibinfo
  {author} {\bibfnamefont {C.}~\bibnamefont {Borgmann}}, \bibinfo {author}
  {\bibfnamefont {M.}~\bibnamefont {Breitenfeldt}}, \bibinfo {author}
  {\bibfnamefont {R.~B.}\ \bibnamefont {Cakirli}}, \bibinfo {author}
  {\bibfnamefont {S.}~\bibnamefont {George}}, \bibinfo {author} {\bibfnamefont
  {F.}~\bibnamefont {Herfurth}}, \bibinfo {author} {\bibfnamefont {J.~D.}\
  \bibnamefont {Holt}}, \bibinfo {author} {\bibfnamefont {M.}~\bibnamefont
  {Kowalska}}, \emph {et~al.},\ }\href {https://doi.org/10.1038/nature12226}
  {\bibfield  {journal} {\bibinfo  {journal} {Nature}\ }\textbf {\bibinfo
  {volume} {498}},\ \bibinfo {pages} {346} (\bibinfo {year}
  {2013})}\BibitemShut {NoStop}%
\bibitem [{\citenamefont {Xu}\ \emph {et~al.}(2015)\citenamefont {Xu},
  \citenamefont {Wang}, \citenamefont {Zhang}, \citenamefont {Xu},
  \citenamefont {Shuai}, \citenamefont {Tu}, \citenamefont {Litvinov},
  \citenamefont {Zhou}, \citenamefont {Sun}, \citenamefont {Yuan} \emph
  {et~al.}}]{Xu2015}%
  \BibitemOpen
  \bibfield  {author} {\bibinfo {author} {\bibfnamefont {X.}~\bibnamefont
  {Xu}}, \bibinfo {author} {\bibfnamefont {M.}~\bibnamefont {Wang}}, \bibinfo
  {author} {\bibfnamefont {Y.-H.}\ \bibnamefont {Zhang}}, \bibinfo {author}
  {\bibfnamefont {H.-S.}\ \bibnamefont {Xu}}, \bibinfo {author} {\bibfnamefont
  {P.}~\bibnamefont {Shuai}}, \bibinfo {author} {\bibfnamefont {X.-L.}\
  \bibnamefont {Tu}}, \bibinfo {author} {\bibfnamefont {Y.~A.}\ \bibnamefont
  {Litvinov}}, \bibinfo {author} {\bibfnamefont {X.-H.}\ \bibnamefont {Zhou}},
  \bibinfo {author} {\bibfnamefont {B.-H.}\ \bibnamefont {Sun}}, \bibinfo
  {author} {\bibfnamefont {Y.-J.}\ \bibnamefont {Yuan}}, \emph {et~al.},\
  }\href {https://doi.org/10.1088/1674-1137/39/10/104001} {\bibfield  {journal}
  {\bibinfo  {journal} {Chinese Physics C}\ }\textbf {\bibinfo {volume} {39}},\
  \bibinfo {pages} {104001} (\bibinfo {year} {2015})}\BibitemShut {NoStop}%
\bibitem [{\citenamefont {Xu}\ \emph {et~al.}(2019)\citenamefont {Xu},
  \citenamefont {Wang}, \citenamefont {Blaum}, \citenamefont {Holt},
  \citenamefont {Litvinov}, \citenamefont {Schwenk}, \citenamefont {Simonis},
  \citenamefont {Stroberg}, \citenamefont {Zhang}, \citenamefont {Xu} \emph
  {et~al.}}]{Xu2019}%
  \BibitemOpen
  \bibfield  {author} {\bibinfo {author} {\bibfnamefont {X.}~\bibnamefont
  {Xu}}, \bibinfo {author} {\bibfnamefont {M.}~\bibnamefont {Wang}}, \bibinfo
  {author} {\bibfnamefont {K.}~\bibnamefont {Blaum}}, \bibinfo {author}
  {\bibfnamefont {J.~D.}\ \bibnamefont {Holt}}, \bibinfo {author}
  {\bibfnamefont {Y.~A.}\ \bibnamefont {Litvinov}}, \bibinfo {author}
  {\bibfnamefont {A.}~\bibnamefont {Schwenk}}, \bibinfo {author} {\bibfnamefont
  {J.}~\bibnamefont {Simonis}}, \bibinfo {author} {\bibfnamefont {S.~R.}\
  \bibnamefont {Stroberg}}, \bibinfo {author} {\bibfnamefont {Y.~H.}\
  \bibnamefont {Zhang}}, \bibinfo {author} {\bibfnamefont {H.~S.}\ \bibnamefont
  {Xu}}, \emph {et~al.},\ }\href {https://doi.org/10.1103/PhysRevC.99.064303}
  {\bibfield  {journal} {\bibinfo  {journal} {Phys. Rev. C}\ }\textbf {\bibinfo
  {volume} {99}},\ \bibinfo {pages} {064303} (\bibinfo {year}
  {2019})}\BibitemShut {NoStop}%
\bibitem [{\citenamefont {Enciu}\ \emph {et~al.}(2022)\citenamefont {Enciu},
  \citenamefont {Liu}, \citenamefont {Obertelli}, \citenamefont {Doornenbal},
  \citenamefont {Nowacki}, \citenamefont {Ogata}, \citenamefont {Poves},
  \citenamefont {Yoshida}, \citenamefont {Achouri}, \citenamefont {Baba} \emph
  {et~al.}}]{Enciu2022}%
  \BibitemOpen
  \bibfield  {author} {\bibinfo {author} {\bibfnamefont {M.}~\bibnamefont
  {Enciu}}, \bibinfo {author} {\bibfnamefont {H.~N.}\ \bibnamefont {Liu}},
  \bibinfo {author} {\bibfnamefont {A.}~\bibnamefont {Obertelli}}, \bibinfo
  {author} {\bibfnamefont {P.}~\bibnamefont {Doornenbal}}, \bibinfo {author}
  {\bibfnamefont {F.}~\bibnamefont {Nowacki}}, \bibinfo {author} {\bibfnamefont
  {K.}~\bibnamefont {Ogata}}, \bibinfo {author} {\bibfnamefont
  {A.}~\bibnamefont {Poves}}, \bibinfo {author} {\bibfnamefont
  {K.}~\bibnamefont {Yoshida}}, \bibinfo {author} {\bibfnamefont {N.~L.}\
  \bibnamefont {Achouri}}, \bibinfo {author} {\bibfnamefont {H.}~\bibnamefont
  {Baba}}, \emph {et~al.},\ }\href
  {https://doi.org/10.1103/PhysRevLett.129.262501} {\bibfield  {journal}
  {\bibinfo  {journal} {Phys. Rev. Lett.}\ }\textbf {\bibinfo {volume} {129}},\
  \bibinfo {pages} {262501} (\bibinfo {year} {2022})}\BibitemShut {NoStop}%
\bibitem [{\citenamefont {Garc\'ia~Ruiz}\ \emph {et~al.}(2016)\citenamefont
  {Garc\'ia~Ruiz}, \citenamefont {Bissell}, \citenamefont {Blaum},
  \citenamefont {Ekstr\"om}, \citenamefont {Fr\"ommgen}, \citenamefont {Hagen},
  \citenamefont {Hammen}, \citenamefont {Hebeler}, \citenamefont {Holt},
  \citenamefont {Jansen} \emph {et~al.}}]{GarciaRuiz2016}%
  \BibitemOpen
  \bibfield  {author} {\bibinfo {author} {\bibfnamefont {R.~F.}\ \bibnamefont
  {Garc\'ia~Ruiz}}, \bibinfo {author} {\bibfnamefont {M.~L.}\ \bibnamefont
  {Bissell}}, \bibinfo {author} {\bibfnamefont {K.}~\bibnamefont {Blaum}},
  \bibinfo {author} {\bibfnamefont {A.}~\bibnamefont {Ekstr\"om}}, \bibinfo
  {author} {\bibfnamefont {N.}~\bibnamefont {Fr\"ommgen}}, \bibinfo {author}
  {\bibfnamefont {G.}~\bibnamefont {Hagen}}, \bibinfo {author} {\bibfnamefont
  {M.}~\bibnamefont {Hammen}}, \bibinfo {author} {\bibfnamefont
  {K.}~\bibnamefont {Hebeler}}, \bibinfo {author} {\bibfnamefont {J.~D.}\
  \bibnamefont {Holt}}, \bibinfo {author} {\bibfnamefont {G.~R.}\ \bibnamefont
  {Jansen}}, \emph {et~al.},\ }\href {https://doi.org/10.1038/nphys3645}
  {\bibfield  {journal} {\bibinfo  {journal} {Nature Physics}\ }\textbf
  {\bibinfo {volume} {12}},\ \bibinfo {pages} {594} (\bibinfo {year}
  {2016})}\BibitemShut {NoStop}%
\bibitem [{\citenamefont {Koszor\'us}\ \emph {et~al.}(2021)\citenamefont
  {Koszor\'us}, \citenamefont {Yang}, \citenamefont {Jiang}, \citenamefont
  {Novario}, \citenamefont {Bai}, \citenamefont {Billowes}, \citenamefont
  {Binnersley}, \citenamefont {Bissell}, \citenamefont {Cocolios},
  \citenamefont {Cooper} \emph {et~al.}}]{Koszorus2021}%
  \BibitemOpen
  \bibfield  {author} {\bibinfo {author} {\bibfnamefont {A.}~\bibnamefont
  {Koszor\'us}}, \bibinfo {author} {\bibfnamefont {X.~F.}\ \bibnamefont
  {Yang}}, \bibinfo {author} {\bibfnamefont {W.~G.}\ \bibnamefont {Jiang}},
  \bibinfo {author} {\bibfnamefont {S.~J.}\ \bibnamefont {Novario}}, \bibinfo
  {author} {\bibfnamefont {S.~W.}\ \bibnamefont {Bai}}, \bibinfo {author}
  {\bibfnamefont {J.}~\bibnamefont {Billowes}}, \bibinfo {author}
  {\bibfnamefont {C.~L.}\ \bibnamefont {Binnersley}}, \bibinfo {author}
  {\bibfnamefont {M.~L.}\ \bibnamefont {Bissell}}, \bibinfo {author}
  {\bibfnamefont {T.~E.}\ \bibnamefont {Cocolios}}, \bibinfo {author}
  {\bibfnamefont {B.~S.}\ \bibnamefont {Cooper}}, \emph {et~al.},\ }\href
  {https://doi.org/10.1038/s41567-020-01136-5} {\bibfield  {journal} {\bibinfo
  {journal} {Nature Physics}\ }\textbf {\bibinfo {volume} {17}},\ \bibinfo
  {pages} {439} (\bibinfo {year} {2021})}\BibitemShut {NoStop}%
\bibitem [{\citenamefont {Leistenschneider}\ \emph {et~al.}(2018)\citenamefont
  {Leistenschneider}, \citenamefont {Reiter}, \citenamefont {Ayet
  San~Andr\'es}, \citenamefont {Kootte}, \citenamefont {Holt}, \citenamefont
  {Navr\'atil}, \citenamefont {Babcock}, \citenamefont {Barbieri},
  \citenamefont {Barquest}, \citenamefont {Bergmann} \emph
  {et~al.}}]{Leistenschneider2018}%
  \BibitemOpen
  \bibfield  {author} {\bibinfo {author} {\bibfnamefont {E.}~\bibnamefont
  {Leistenschneider}}, \bibinfo {author} {\bibfnamefont {M.~P.}\ \bibnamefont
  {Reiter}}, \bibinfo {author} {\bibfnamefont {S.}~\bibnamefont {Ayet
  San~Andr\'es}}, \bibinfo {author} {\bibfnamefont {B.}~\bibnamefont {Kootte}},
  \bibinfo {author} {\bibfnamefont {J.~D.}\ \bibnamefont {Holt}}, \bibinfo
  {author} {\bibfnamefont {P.}~\bibnamefont {Navr\'atil}}, \bibinfo {author}
  {\bibfnamefont {C.}~\bibnamefont {Babcock}}, \bibinfo {author} {\bibfnamefont
  {C.}~\bibnamefont {Barbieri}}, \bibinfo {author} {\bibfnamefont {B.~R.}\
  \bibnamefont {Barquest}}, \bibinfo {author} {\bibfnamefont {J.}~\bibnamefont
  {Bergmann}}, \emph {et~al.},\ }\href
  {https://doi.org/10.1103/PhysRevLett.120.062503} {\bibfield  {journal}
  {\bibinfo  {journal} {Phys. Rev. Lett.}\ }\textbf {\bibinfo {volume} {120}},\
  \bibinfo {pages} {062503} (\bibinfo {year} {2018})}\BibitemShut {NoStop}%
\bibitem [{\citenamefont {Steppenbeck}\ \emph {et~al.}(2015)\citenamefont
  {Steppenbeck}, \citenamefont {Takeuchi}, \citenamefont {Aoi}, \citenamefont
  {Doornenbal}, \citenamefont {Matsushita}, \citenamefont {Wang}, \citenamefont
  {Utsuno}, \citenamefont {Baba}, \citenamefont {Go}, \citenamefont {Lee} \emph
  {et~al.}}]{Steppenbeck2015}%
  \BibitemOpen
  \bibfield  {author} {\bibinfo {author} {\bibfnamefont {D.}~\bibnamefont
  {Steppenbeck}}, \bibinfo {author} {\bibfnamefont {S.}~\bibnamefont
  {Takeuchi}}, \bibinfo {author} {\bibfnamefont {N.}~\bibnamefont {Aoi}},
  \bibinfo {author} {\bibfnamefont {P.}~\bibnamefont {Doornenbal}}, \bibinfo
  {author} {\bibfnamefont {M.}~\bibnamefont {Matsushita}}, \bibinfo {author}
  {\bibfnamefont {H.}~\bibnamefont {Wang}}, \bibinfo {author} {\bibfnamefont
  {Y.}~\bibnamefont {Utsuno}}, \bibinfo {author} {\bibfnamefont
  {H.}~\bibnamefont {Baba}}, \bibinfo {author} {\bibfnamefont {S.}~\bibnamefont
  {Go}}, \bibinfo {author} {\bibfnamefont {J.}~\bibnamefont {Lee}}, \emph
  {et~al.},\ }\href {https://doi.org/10.1103/PhysRevLett.114.252501} {\bibfield
   {journal} {\bibinfo  {journal} {Phys. Rev. Lett.}\ }\textbf {\bibinfo
  {volume} {114}},\ \bibinfo {pages} {252501} (\bibinfo {year}
  {2015})}\BibitemShut {NoStop}%
\bibitem [{\citenamefont {Cort\'es}\ \emph {et~al.}(2020)\citenamefont
  {Cort\'es}, \citenamefont {Rodriguez}, \citenamefont {Doornenbal},
  \citenamefont {Obertelli}, \citenamefont {Holt}, \citenamefont {Men\'endez},
  \citenamefont {Ogata}, \citenamefont {Schwenk}, \citenamefont {Shimizu},
  \citenamefont {Simonis} \emph {et~al.}}]{Cortes2020}%
  \BibitemOpen
  \bibfield  {author} {\bibinfo {author} {\bibfnamefont {M.~L.}\ \bibnamefont
  {Cort\'es}}, \bibinfo {author} {\bibfnamefont {W.}~\bibnamefont {Rodriguez}},
  \bibinfo {author} {\bibfnamefont {P.}~\bibnamefont {Doornenbal}}, \bibinfo
  {author} {\bibfnamefont {A.}~\bibnamefont {Obertelli}}, \bibinfo {author}
  {\bibfnamefont {J.~D.}\ \bibnamefont {Holt}}, \bibinfo {author}
  {\bibfnamefont {J.}~\bibnamefont {Men\'endez}}, \bibinfo {author}
  {\bibfnamefont {K.}~\bibnamefont {Ogata}}, \bibinfo {author} {\bibfnamefont
  {A.}~\bibnamefont {Schwenk}}, \bibinfo {author} {\bibfnamefont
  {N.}~\bibnamefont {Shimizu}}, \bibinfo {author} {\bibfnamefont
  {J.}~\bibnamefont {Simonis}}, \emph {et~al.},\ }\href
  {https://doi.org/10.1103/PhysRevC.102.064320} {\bibfield  {journal} {\bibinfo
   {journal} {Phys. Rev. C}\ }\textbf {\bibinfo {volume} {102}},\ \bibinfo
  {pages} {064320} (\bibinfo {year} {2020})}\BibitemShut {NoStop}%
\bibitem [{\citenamefont {Sahoo}\ and\ \citenamefont
  {Srivastava}(2025)}]{Sahoo2025}%
  \BibitemOpen
  \bibfield  {author} {\bibinfo {author} {\bibfnamefont {S.}~\bibnamefont
  {Sahoo}}\ and\ \bibinfo {author} {\bibfnamefont {P.~C.}\ \bibnamefont
  {Srivastava}},\ }\href {https://doi.org/10.1103/423y-znv8} {\bibfield
  {journal} {\bibinfo  {journal} {Phys. Rev. C}\ }\textbf {\bibinfo {volume}
  {112}},\ \bibinfo {pages} {L021301} (\bibinfo {year} {2025})}\BibitemShut
  {NoStop}%
\bibitem [{\citenamefont {Li}\ \emph {et~al.}(2016)\citenamefont {Li},
  \citenamefont {Margueron}, \citenamefont {Long},\ and\ \citenamefont
  {Giai}}]{Li2016}%
  \BibitemOpen
  \bibfield  {author} {\bibinfo {author} {\bibfnamefont {J.~J.}\ \bibnamefont
  {Li}}, \bibinfo {author} {\bibfnamefont {J.}~\bibnamefont {Margueron}},
  \bibinfo {author} {\bibfnamefont {W.~H.}\ \bibnamefont {Long}},\ and\
  \bibinfo {author} {\bibfnamefont {N.~V.}\ \bibnamefont {Giai}},\ }\href
  {https://doi.org/10.1016/j.physletb.2015.12.004} {\bibfield  {journal}
  {\bibinfo  {journal} {Physics Letters B}\ }\textbf {\bibinfo {volume}
  {753}},\ \bibinfo {pages} {97} (\bibinfo {year} {2016})}\BibitemShut
  {NoStop}%
\bibitem [{\citenamefont {Liu}\ \emph {et~al.}(2020)\citenamefont {Liu},
  \citenamefont {Niu},\ and\ \citenamefont {Long}}]{Liu2020}%
  \BibitemOpen
  \bibfield  {author} {\bibinfo {author} {\bibfnamefont {J.}~\bibnamefont
  {Liu}}, \bibinfo {author} {\bibfnamefont {Y.~F.}\ \bibnamefont {Niu}},\ and\
  \bibinfo {author} {\bibfnamefont {W.~H.}\ \bibnamefont {Long}},\ }\href
  {https://doi.org/10.1016/j.physletb.2020.135524} {\bibfield  {journal}
  {\bibinfo  {journal} {Physics Letters B}\ }\textbf {\bibinfo {volume}
  {806}},\ \bibinfo {pages} {135524} (\bibinfo {year} {2020})}\BibitemShut
  {NoStop}%
\bibitem [{\citenamefont {Grasso}(2014)}]{Grasso2014}%
  \BibitemOpen
  \bibfield  {author} {\bibinfo {author} {\bibfnamefont {M.}~\bibnamefont
  {Grasso}},\ }\href {https://doi.org/10.1103/PhysRevC.89.034316} {\bibfield
  {journal} {\bibinfo  {journal} {Phys. Rev. C}\ }\textbf {\bibinfo {volume}
  {89}},\ \bibinfo {pages} {034316} (\bibinfo {year} {2014})}\BibitemShut
  {NoStop}%
\bibitem [{\citenamefont {Honma}\ \emph {et~al.}(2002)\citenamefont {Honma},
  \citenamefont {Otsuka}, \citenamefont {Brown},\ and\ \citenamefont
  {Mizusaki}}]{Honma2002}%
  \BibitemOpen
  \bibfield  {author} {\bibinfo {author} {\bibfnamefont {M.}~\bibnamefont
  {Honma}}, \bibinfo {author} {\bibfnamefont {T.}~\bibnamefont {Otsuka}},
  \bibinfo {author} {\bibfnamefont {B.~A.}\ \bibnamefont {Brown}},\ and\
  \bibinfo {author} {\bibfnamefont {T.}~\bibnamefont {Mizusaki}},\ }\href
  {https://doi.org/10.1103/PhysRevC.65.061301} {\bibfield  {journal} {\bibinfo
  {journal} {Phys. Rev. C}\ }\textbf {\bibinfo {volume} {65}},\ \bibinfo
  {pages} {061301} (\bibinfo {year} {2002})}\BibitemShut {NoStop}%
\bibitem [{\citenamefont {Mantica}\ \emph {et~al.}(2003)\citenamefont
  {Mantica}, \citenamefont {Morton}, \citenamefont {Brown}, \citenamefont
  {Davies}, \citenamefont {Glasmacher}, \citenamefont {Groh}, \citenamefont
  {Liddick}, \citenamefont {Morrissey}, \citenamefont {Mueller}, \citenamefont
  {Schatz}, \citenamefont {Stolz}, \citenamefont {Tabor}, \citenamefont
  {Honma}, \citenamefont {Horoi},\ and\ \citenamefont {Otsuka}}]{Mantica2003}%
  \BibitemOpen
  \bibfield  {author} {\bibinfo {author} {\bibfnamefont {P.~F.}\ \bibnamefont
  {Mantica}}, \bibinfo {author} {\bibfnamefont {A.~C.}\ \bibnamefont {Morton}},
  \bibinfo {author} {\bibfnamefont {B.~A.}\ \bibnamefont {Brown}}, \bibinfo
  {author} {\bibfnamefont {A.~D.}\ \bibnamefont {Davies}}, \bibinfo {author}
  {\bibfnamefont {T.}~\bibnamefont {Glasmacher}}, \bibinfo {author}
  {\bibfnamefont {D.~E.}\ \bibnamefont {Groh}}, \bibinfo {author}
  {\bibfnamefont {S.~N.}\ \bibnamefont {Liddick}}, \bibinfo {author}
  {\bibfnamefont {D.~J.}\ \bibnamefont {Morrissey}}, \bibinfo {author}
  {\bibfnamefont {W.~F.}\ \bibnamefont {Mueller}}, \bibinfo {author}
  {\bibfnamefont {H.}~\bibnamefont {Schatz}}, \bibinfo {author} {\bibfnamefont
  {A.}~\bibnamefont {Stolz}}, \bibinfo {author} {\bibfnamefont {S.~L.}\
  \bibnamefont {Tabor}}, \bibinfo {author} {\bibfnamefont {M.}~\bibnamefont
  {Honma}}, \bibinfo {author} {\bibfnamefont {M.}~\bibnamefont {Horoi}},\ and\
  \bibinfo {author} {\bibfnamefont {T.}~\bibnamefont {Otsuka}},\ }\href
  {https://doi.org/10.1103/PhysRevC.67.014311} {\bibfield  {journal} {\bibinfo
  {journal} {Phys. Rev. C}\ }\textbf {\bibinfo {volume} {67}},\ \bibinfo
  {pages} {014311} (\bibinfo {year} {2003})}\BibitemShut {NoStop}%
\bibitem [{\citenamefont {Liddick}\ \emph {et~al.}(2004)\citenamefont
  {Liddick}, \citenamefont {Mantica}, \citenamefont {Janssens}, \citenamefont
  {Broda}, \citenamefont {Brown}, \citenamefont {Carpenter}, \citenamefont
  {Fornal}, \citenamefont {Honma}, \citenamefont {Mizusaki}, \citenamefont
  {Morton} \emph {et~al.}}]{Liddick2004}%
  \BibitemOpen
  \bibfield  {author} {\bibinfo {author} {\bibfnamefont {S.~N.}\ \bibnamefont
  {Liddick}}, \bibinfo {author} {\bibfnamefont {P.~F.}\ \bibnamefont
  {Mantica}}, \bibinfo {author} {\bibfnamefont {R.~V.~F.}\ \bibnamefont
  {Janssens}}, \bibinfo {author} {\bibfnamefont {R.}~\bibnamefont {Broda}},
  \bibinfo {author} {\bibfnamefont {B.~A.}\ \bibnamefont {Brown}}, \bibinfo
  {author} {\bibfnamefont {M.~P.}\ \bibnamefont {Carpenter}}, \bibinfo {author}
  {\bibfnamefont {B.}~\bibnamefont {Fornal}}, \bibinfo {author} {\bibfnamefont
  {M.}~\bibnamefont {Honma}}, \bibinfo {author} {\bibfnamefont
  {T.}~\bibnamefont {Mizusaki}}, \bibinfo {author} {\bibfnamefont {A.~C.}\
  \bibnamefont {Morton}}, \emph {et~al.},\ }\href
  {https://doi.org/10.1103/PhysRevLett.92.072502} {\bibfield  {journal}
  {\bibinfo  {journal} {Phys. Rev. Lett.}\ }\textbf {\bibinfo {volume} {92}},\
  \bibinfo {pages} {072502} (\bibinfo {year} {2004})}\BibitemShut {NoStop}%
\bibitem [{\citenamefont {Honma}\ \emph {et~al.}(2005)\citenamefont {Honma},
  \citenamefont {Otsuka}, \citenamefont {Brown},\ and\ \citenamefont
  {Mizusaki}}]{Honma2005}%
  \BibitemOpen
  \bibfield  {author} {\bibinfo {author} {\bibfnamefont {M.}~\bibnamefont
  {Honma}}, \bibinfo {author} {\bibfnamefont {T.}~\bibnamefont {Otsuka}},
  \bibinfo {author} {\bibfnamefont {B.~A.}\ \bibnamefont {Brown}},\ and\
  \bibinfo {author} {\bibfnamefont {T.}~\bibnamefont {Mizusaki}},\ }\href
  {https://doi.org/10.1140/epjad/i2005-06-032-2} {\bibfield  {journal}
  {\bibinfo  {journal} {The European Physical Journal A - Hadrons and Nuclei}\
  }\textbf {\bibinfo {volume} {25}},\ \bibinfo {pages} {499} (\bibinfo {year}
  {2005})}\BibitemShut {NoStop}%
\bibitem [{\citenamefont {Coraggio}\ \emph {et~al.}(2009)\citenamefont
  {Coraggio}, \citenamefont {Covello}, \citenamefont {Gargano},\ and\
  \citenamefont {Itaco}}]{Coraggio2009}%
  \BibitemOpen
  \bibfield  {author} {\bibinfo {author} {\bibfnamefont {L.}~\bibnamefont
  {Coraggio}}, \bibinfo {author} {\bibfnamefont {A.}~\bibnamefont {Covello}},
  \bibinfo {author} {\bibfnamefont {A.}~\bibnamefont {Gargano}},\ and\ \bibinfo
  {author} {\bibfnamefont {N.}~\bibnamefont {Itaco}},\ }\href
  {https://doi.org/10.1103/PhysRevC.80.044311} {\bibfield  {journal} {\bibinfo
  {journal} {Phys. Rev. C}\ }\textbf {\bibinfo {volume} {80}},\ \bibinfo
  {pages} {044311} (\bibinfo {year} {2009})}\BibitemShut {NoStop}%
\bibitem [{\citenamefont {Rodr\'{\i}guez}\ and\ \citenamefont
  {Egido}(2007)}]{Rodriguez2007}%
  \BibitemOpen
  \bibfield  {author} {\bibinfo {author} {\bibfnamefont {T.~R.}\ \bibnamefont
  {Rodr\'{\i}guez}}\ and\ \bibinfo {author} {\bibfnamefont {J.~L.}\
  \bibnamefont {Egido}},\ }\href
  {https://doi.org/10.1103/PhysRevLett.99.062501} {\bibfield  {journal}
  {\bibinfo  {journal} {Phys. Rev. Lett.}\ }\textbf {\bibinfo {volume} {99}},\
  \bibinfo {pages} {062501} (\bibinfo {year} {2007})}\BibitemShut {NoStop}%
\bibitem [{\citenamefont {Hagen}\ \emph {et~al.}(2012)\citenamefont {Hagen},
  \citenamefont {Hjorth-Jensen}, \citenamefont {Jansen}, \citenamefont
  {Machleidt},\ and\ \citenamefont {Papenbrock}}]{Hagen2012}%
  \BibitemOpen
  \bibfield  {author} {\bibinfo {author} {\bibfnamefont {G.}~\bibnamefont
  {Hagen}}, \bibinfo {author} {\bibfnamefont {M.}~\bibnamefont
  {Hjorth-Jensen}}, \bibinfo {author} {\bibfnamefont {G.~R.}\ \bibnamefont
  {Jansen}}, \bibinfo {author} {\bibfnamefont {R.}~\bibnamefont {Machleidt}},\
  and\ \bibinfo {author} {\bibfnamefont {T.}~\bibnamefont {Papenbrock}},\
  }\href {https://doi.org/10.1103/PhysRevLett.109.032502} {\bibfield  {journal}
  {\bibinfo  {journal} {Phys. Rev. Lett.}\ }\textbf {\bibinfo {volume} {109}},\
  \bibinfo {pages} {032502} (\bibinfo {year} {2012})}\BibitemShut {NoStop}%
\bibitem [{\citenamefont {Liu}(2024)}]{Liu2024}%
  \BibitemOpen
  \bibfield  {author} {\bibinfo {author} {\bibfnamefont {J.}~\bibnamefont
  {Liu}},\ }\emph {\bibinfo {title} {Configuration interactions based on
  relativistic Hartree-Fock theory}},\ \href@noop {} {Ph.D. thesis},\ \bibinfo
  {school} {Lanzhou University}, \bibinfo {address} {Lanzhou, China} (\bibinfo
  {year} {2024})\BibitemShut {NoStop}%
\bibitem [{\citenamefont {Zhang}\ \emph {et~al.}(2007)\citenamefont {Zhang},
  \citenamefont {Ren}, \citenamefont {Zhi},\ and\ \citenamefont
  {Zheng}}]{Zhang2007}%
  \BibitemOpen
  \bibfield  {author} {\bibinfo {author} {\bibfnamefont {X.}~\bibnamefont
  {Zhang}}, \bibinfo {author} {\bibfnamefont {Z.}~\bibnamefont {Ren}}, \bibinfo
  {author} {\bibfnamefont {Q.}~\bibnamefont {Zhi}},\ and\ \bibinfo {author}
  {\bibfnamefont {Q.}~\bibnamefont {Zheng}},\ }\href
  {https://doi.org/10.1088/0954-3899/34/12/007} {\bibfield  {journal} {\bibinfo
   {journal} {Journal of Physics G: Nuclear and Particle Physics}\ }\textbf
  {\bibinfo {volume} {34}},\ \bibinfo {pages} {2611} (\bibinfo {year}
  {2007})}\BibitemShut {NoStop}%
\bibitem [{\citenamefont {Sorlin}\ \emph {et~al.}(1993)\citenamefont {Sorlin},
  \citenamefont {Guillemaud-Mueller}, \citenamefont {Mueller}, \citenamefont
  {Borrel}, \citenamefont {Dogny}, \citenamefont {Pougheon}, \citenamefont
  {Kratz}, \citenamefont {Gabelmann}, \citenamefont {Pfeiffer}, \citenamefont
  {W\"ohr} \emph {et~al.}}]{Sorlin1993}%
  \BibitemOpen
  \bibfield  {author} {\bibinfo {author} {\bibfnamefont {O.}~\bibnamefont
  {Sorlin}}, \bibinfo {author} {\bibfnamefont {D.}~\bibnamefont
  {Guillemaud-Mueller}}, \bibinfo {author} {\bibfnamefont {A.~C.}\ \bibnamefont
  {Mueller}}, \bibinfo {author} {\bibfnamefont {V.}~\bibnamefont {Borrel}},
  \bibinfo {author} {\bibfnamefont {S.}~\bibnamefont {Dogny}}, \bibinfo
  {author} {\bibfnamefont {F.}~\bibnamefont {Pougheon}}, \bibinfo {author}
  {\bibfnamefont {K.-L.}\ \bibnamefont {Kratz}}, \bibinfo {author}
  {\bibfnamefont {H.}~\bibnamefont {Gabelmann}}, \bibinfo {author}
  {\bibfnamefont {B.}~\bibnamefont {Pfeiffer}}, \bibinfo {author}
  {\bibfnamefont {A.}~\bibnamefont {W\"ohr}}, \emph {et~al.},\ }\href
  {https://doi.org/10.1103/PhysRevC.47.2941} {\bibfield  {journal} {\bibinfo
  {journal} {Phys. Rev. C}\ }\textbf {\bibinfo {volume} {47}},\ \bibinfo
  {pages} {2941} (\bibinfo {year} {1993})}\BibitemShut {NoStop}%
\bibitem [{\citenamefont {Yoshida}\ \emph {et~al.}(2023)\citenamefont
  {Yoshida}, \citenamefont {Niu},\ and\ \citenamefont {Minato}}]{Yoshida2023}%
  \BibitemOpen
  \bibfield  {author} {\bibinfo {author} {\bibfnamefont {K.}~\bibnamefont
  {Yoshida}}, \bibinfo {author} {\bibfnamefont {Y.}~\bibnamefont {Niu}},\ and\
  \bibinfo {author} {\bibfnamefont {F.}~\bibnamefont {Minato}},\ }\href
  {https://doi.org/10.1103/PhysRevC.108.034305} {\bibfield  {journal} {\bibinfo
   {journal} {Phys. Rev. C}\ }\textbf {\bibinfo {volume} {108}},\ \bibinfo
  {pages} {034305} (\bibinfo {year} {2023})}\BibitemShut {NoStop}%
\bibitem [{\citenamefont {Yoshida}(2019)}]{Yoshida2019}%
  \BibitemOpen
  \bibfield  {author} {\bibinfo {author} {\bibfnamefont {K.}~\bibnamefont
  {Yoshida}},\ }\href {https://doi.org/10.1103/PhysRevC.100.024316} {\bibfield
  {journal} {\bibinfo  {journal} {Phys. Rev. C}\ }\textbf {\bibinfo {volume}
  {100}},\ \bibinfo {pages} {024316} (\bibinfo {year} {2019})}\BibitemShut
  {NoStop}%
\bibitem [{\citenamefont {Xu}\ \emph {et~al.}(2014)\citenamefont {Xu},
  \citenamefont {Nishimura}, \citenamefont {Lorusso}, \citenamefont {Browne},
  \citenamefont {Doornenbal}, \citenamefont {Gey}, \citenamefont {Jung},
  \citenamefont {Li}, \citenamefont {Niikura}, \citenamefont {S\"oderstr\"om}
  \emph {et~al.}}]{Xu2014}%
  \BibitemOpen
  \bibfield  {author} {\bibinfo {author} {\bibfnamefont {Z.~Y.}\ \bibnamefont
  {Xu}}, \bibinfo {author} {\bibfnamefont {S.}~\bibnamefont {Nishimura}},
  \bibinfo {author} {\bibfnamefont {G.}~\bibnamefont {Lorusso}}, \bibinfo
  {author} {\bibfnamefont {F.}~\bibnamefont {Browne}}, \bibinfo {author}
  {\bibfnamefont {P.}~\bibnamefont {Doornenbal}}, \bibinfo {author}
  {\bibfnamefont {G.}~\bibnamefont {Gey}}, \bibinfo {author} {\bibfnamefont
  {H.-S.}\ \bibnamefont {Jung}}, \bibinfo {author} {\bibfnamefont
  {Z.}~\bibnamefont {Li}}, \bibinfo {author} {\bibfnamefont {M.}~\bibnamefont
  {Niikura}}, \bibinfo {author} {\bibfnamefont {P.-A.}\ \bibnamefont
  {S\"oderstr\"om}}, \emph {et~al.},\ }\href
  {https://doi.org/10.1103/PhysRevLett.113.032505} {\bibfield  {journal}
  {\bibinfo  {journal} {Phys. Rev. Lett.}\ }\textbf {\bibinfo {volume} {113}},\
  \bibinfo {pages} {032505} (\bibinfo {year} {2014})}\BibitemShut {NoStop}%
\bibitem [{\citenamefont {Zeng}\ \emph {et~al.}(2025)\citenamefont {Zeng} \emph
  {et~al.}}]{QBZeng2025}%
  \BibitemOpen
  \bibfield  {author} {\bibinfo {author} {\bibfnamefont {Q.}~\bibnamefont
  {Zeng}} \emph {et~al.},\ }\href {https://doi.org/10.1103/227j-q7zf}
  {\bibfield  {journal} {\bibinfo  {journal} {Phys. Rev. Lett.}\ ,\ } (\bibinfo
  {year} {2025})},\ \bibinfo {note} {accepted for publication}\BibitemShut
  {NoStop}%
\bibitem [{\citenamefont {Engel}\ \emph {et~al.}(1999)\citenamefont {Engel},
  \citenamefont {Bender}, \citenamefont {Dobaczewski}, \citenamefont
  {Nazarewicz},\ and\ \citenamefont {Surman}}]{Engel1999}%
  \BibitemOpen
  \bibfield  {author} {\bibinfo {author} {\bibfnamefont {J.}~\bibnamefont
  {Engel}}, \bibinfo {author} {\bibfnamefont {M.}~\bibnamefont {Bender}},
  \bibinfo {author} {\bibfnamefont {J.}~\bibnamefont {Dobaczewski}}, \bibinfo
  {author} {\bibfnamefont {W.}~\bibnamefont {Nazarewicz}},\ and\ \bibinfo
  {author} {\bibfnamefont {R.}~\bibnamefont {Surman}},\ }\href
  {https://doi.org/10.1103/PhysRevC.60.014302} {\bibfield  {journal} {\bibinfo
  {journal} {Phys. Rev. C}\ }\textbf {\bibinfo {volume} {60}},\ \bibinfo
  {pages} {014302} (\bibinfo {year} {1999})}\BibitemShut {NoStop}%
\bibitem [{\citenamefont {Mustonen}\ and\ \citenamefont
  {Engel}(2016)}]{Mustonen2016}%
  \BibitemOpen
  \bibfield  {author} {\bibinfo {author} {\bibfnamefont {M.~T.}\ \bibnamefont
  {Mustonen}}\ and\ \bibinfo {author} {\bibfnamefont {J.}~\bibnamefont
  {Engel}},\ }\href {https://doi.org/10.1103/PhysRevC.93.014304} {\bibfield
  {journal} {\bibinfo  {journal} {Phys. Rev. C}\ }\textbf {\bibinfo {volume}
  {93}},\ \bibinfo {pages} {014304} (\bibinfo {year} {2016})}\BibitemShut
  {NoStop}%
\bibitem [{\citenamefont {Minato}\ \emph {et~al.}(2022)\citenamefont {Minato},
  \citenamefont {Niu},\ and\ \citenamefont {Liang}}]{Minato2022}%
  \BibitemOpen
  \bibfield  {author} {\bibinfo {author} {\bibfnamefont {F.}~\bibnamefont
  {Minato}}, \bibinfo {author} {\bibfnamefont {Z.}~\bibnamefont {Niu}},\ and\
  \bibinfo {author} {\bibfnamefont {H.}~\bibnamefont {Liang}},\ }\href
  {https://doi.org/10.1103/PhysRevC.106.024306} {\bibfield  {journal} {\bibinfo
   {journal} {Phys. Rev. C}\ }\textbf {\bibinfo {volume} {106}},\ \bibinfo
  {pages} {024306} (\bibinfo {year} {2022})}\BibitemShut {NoStop}%
\bibitem [{\citenamefont {Nik\ifmmode \check{s}\else
  \v{s}\fi{}i\ifmmode~\acute{c}\else \'{c}\fi{}}\ \emph
  {et~al.}(2005)\citenamefont {Nik\ifmmode \check{s}\else
  \v{s}\fi{}i\ifmmode~\acute{c}\else \'{c}\fi{}}, \citenamefont {Marketin},
  \citenamefont {Vretenar}, \citenamefont {Paar},\ and\ \citenamefont
  {Ring}}]{Niksic2005}%
  \BibitemOpen
  \bibfield  {author} {\bibinfo {author} {\bibfnamefont {T.}~\bibnamefont
  {Nik\ifmmode \check{s}\else \v{s}\fi{}i\ifmmode~\acute{c}\else \'{c}\fi{}}},
  \bibinfo {author} {\bibfnamefont {T.}~\bibnamefont {Marketin}}, \bibinfo
  {author} {\bibfnamefont {D.}~\bibnamefont {Vretenar}}, \bibinfo {author}
  {\bibfnamefont {N.}~\bibnamefont {Paar}},\ and\ \bibinfo {author}
  {\bibfnamefont {P.}~\bibnamefont {Ring}},\ }\href
  {https://doi.org/10.1103/PhysRevC.71.014308} {\bibfield  {journal} {\bibinfo
  {journal} {Phys. Rev. C}\ }\textbf {\bibinfo {volume} {71}},\ \bibinfo
  {pages} {014308} (\bibinfo {year} {2005})}\BibitemShut {NoStop}%
\bibitem [{\citenamefont {Niu}\ \emph {et~al.}(2013)\citenamefont {Niu},
  \citenamefont {Niu}, \citenamefont {Liang}, \citenamefont {Long},
  \citenamefont {Nik\v{s}i\'c}, \citenamefont {Vretenar},\ and\ \citenamefont
  {Meng}}]{Niu2013}%
  \BibitemOpen
  \bibfield  {author} {\bibinfo {author} {\bibfnamefont {Z.~M.}\ \bibnamefont
  {Niu}}, \bibinfo {author} {\bibfnamefont {Y.~F.}\ \bibnamefont {Niu}},
  \bibinfo {author} {\bibfnamefont {H.~Z.}\ \bibnamefont {Liang}}, \bibinfo
  {author} {\bibfnamefont {W.~H.}\ \bibnamefont {Long}}, \bibinfo {author}
  {\bibfnamefont {T.}~\bibnamefont {Nik\v{s}i\'c}}, \bibinfo {author}
  {\bibfnamefont {D.}~\bibnamefont {Vretenar}},\ and\ \bibinfo {author}
  {\bibfnamefont {J.}~\bibnamefont {Meng}},\ }\href
  {https://doi.org/10.1016/j.physletb.2013.04.048} {\bibfield  {journal}
  {\bibinfo  {journal} {Physics Letters B}\ }\textbf {\bibinfo {volume}
  {723}},\ \bibinfo {pages} {172} (\bibinfo {year} {2013})}\BibitemShut
  {NoStop}%
\bibitem [{\citenamefont {Marketin}\ \emph {et~al.}(2016)\citenamefont
  {Marketin}, \citenamefont {Huther},\ and\ \citenamefont
  {Mart\'{\i}nez-Pinedo}}]{Marketin2016}%
  \BibitemOpen
  \bibfield  {author} {\bibinfo {author} {\bibfnamefont {T.}~\bibnamefont
  {Marketin}}, \bibinfo {author} {\bibfnamefont {L.}~\bibnamefont {Huther}},\
  and\ \bibinfo {author} {\bibfnamefont {G.}~\bibnamefont
  {Mart\'{\i}nez-Pinedo}},\ }\href {https://doi.org/10.1103/PhysRevC.93.025805}
  {\bibfield  {journal} {\bibinfo  {journal} {Phys. Rev. C}\ }\textbf {\bibinfo
  {volume} {93}},\ \bibinfo {pages} {025805} (\bibinfo {year}
  {2016})}\BibitemShut {NoStop}%
\bibitem [{\citenamefont {Long}\ \emph {et~al.}(2007)\citenamefont {Long},
  \citenamefont {Sagawa}, \citenamefont {Giai},\ and\ \citenamefont
  {Meng}}]{Long2007}%
  \BibitemOpen
  \bibfield  {author} {\bibinfo {author} {\bibfnamefont {W.}~\bibnamefont
  {Long}}, \bibinfo {author} {\bibfnamefont {H.}~\bibnamefont {Sagawa}},
  \bibinfo {author} {\bibfnamefont {N.~V.}\ \bibnamefont {Giai}},\ and\
  \bibinfo {author} {\bibfnamefont {J.}~\bibnamefont {Meng}},\ }\href
  {https://doi.org/10.1103/PhysRevC.76.034314} {\bibfield  {journal} {\bibinfo
  {journal} {Phys. Rev. C}\ }\textbf {\bibinfo {volume} {76}},\ \bibinfo
  {pages} {034314} (\bibinfo {year} {2007})}\BibitemShut {NoStop}%
\bibitem [{\citenamefont {Long}\ \emph {et~al.}(2010)\citenamefont {Long},
  \citenamefont {Ring}, \citenamefont {Giai},\ and\ \citenamefont
  {Meng}}]{Long2010}%
  \BibitemOpen
  \bibfield  {author} {\bibinfo {author} {\bibfnamefont {W.~H.}\ \bibnamefont
  {Long}}, \bibinfo {author} {\bibfnamefont {P.}~\bibnamefont {Ring}}, \bibinfo
  {author} {\bibfnamefont {N.~V.}\ \bibnamefont {Giai}},\ and\ \bibinfo
  {author} {\bibfnamefont {J.}~\bibnamefont {Meng}},\ }\href
  {https://doi.org/10.1103/PhysRevC.81.024308} {\bibfield  {journal} {\bibinfo
  {journal} {Phys. Rev. C}\ }\textbf {\bibinfo {volume} {81}},\ \bibinfo
  {pages} {024308} (\bibinfo {year} {2010})}\BibitemShut {NoStop}%
\bibitem [{\citenamefont {Geng}\ and\ \citenamefont {Long}(2022)}]{Geng2022}%
  \BibitemOpen
  \bibfield  {author} {\bibinfo {author} {\bibfnamefont {J.}~\bibnamefont
  {Geng}}\ and\ \bibinfo {author} {\bibfnamefont {W.~H.}\ \bibnamefont
  {Long}},\ }\href {https://doi.org/10.1103/PhysRevC.105.034329} {\bibfield
  {journal} {\bibinfo  {journal} {Phys. Rev. C}\ }\textbf {\bibinfo {volume}
  {105}},\ \bibinfo {pages} {034329} (\bibinfo {year} {2022})}\BibitemShut
  {NoStop}%
\bibitem [{\citenamefont {Geng}\ \emph {et~al.}(2024)\citenamefont {Geng},
  \citenamefont {Zhao}, \citenamefont {Niu},\ and\ \citenamefont
  {Long}}]{Geng2024}%
  \BibitemOpen
  \bibfield  {author} {\bibinfo {author} {\bibfnamefont {J.}~\bibnamefont
  {Geng}}, \bibinfo {author} {\bibfnamefont {P.~W.}\ \bibnamefont {Zhao}},
  \bibinfo {author} {\bibfnamefont {Y.~F.}\ \bibnamefont {Niu}},\ and\ \bibinfo
  {author} {\bibfnamefont {W.~H.}\ \bibnamefont {Long}},\ }\href
  {https://doi.org/https://doi.org/10.1016/j.physletb.2024.139036} {\bibfield
  {journal} {\bibinfo  {journal} {Physics Letters B}\ }\textbf {\bibinfo
  {volume} {858}},\ \bibinfo {pages} {139036} (\bibinfo {year}
  {2024})}\BibitemShut {NoStop}%
\bibitem [{\citenamefont {Paar}\ \emph {et~al.}(2004)\citenamefont {Paar},
  \citenamefont {Nik\ifmmode \check{s}\else \v{s}\fi{}i\ifmmode~\acute{c}\else
  \'{c}\fi{}}, \citenamefont {Vretenar},\ and\ \citenamefont
  {Ring}}]{Paar2004}%
  \BibitemOpen
  \bibfield  {author} {\bibinfo {author} {\bibfnamefont {N.}~\bibnamefont
  {Paar}}, \bibinfo {author} {\bibfnamefont {T.}~\bibnamefont {Nik\ifmmode
  \check{s}\else \v{s}\fi{}i\ifmmode~\acute{c}\else \'{c}\fi{}}}, \bibinfo
  {author} {\bibfnamefont {D.}~\bibnamefont {Vretenar}},\ and\ \bibinfo
  {author} {\bibfnamefont {P.}~\bibnamefont {Ring}},\ }\href
  {https://doi.org/10.1103/PhysRevC.69.054303} {\bibfield  {journal} {\bibinfo
  {journal} {Phys. Rev. C}\ }\textbf {\bibinfo {volume} {69}},\ \bibinfo
  {pages} {054303} (\bibinfo {year} {2004})}\BibitemShut {NoStop}%
\bibitem [{\citenamefont {Niu}\ \emph {et~al.}(2017)\citenamefont {Niu},
  \citenamefont {Niu}, \citenamefont {Liang}, \citenamefont {Long},\ and\
  \citenamefont {Meng}}]{Niu2017}%
  \BibitemOpen
  \bibfield  {author} {\bibinfo {author} {\bibfnamefont {Z.~M.}\ \bibnamefont
  {Niu}}, \bibinfo {author} {\bibfnamefont {Y.~F.}\ \bibnamefont {Niu}},
  \bibinfo {author} {\bibfnamefont {H.~Z.}\ \bibnamefont {Liang}}, \bibinfo
  {author} {\bibfnamefont {W.~H.}\ \bibnamefont {Long}},\ and\ \bibinfo
  {author} {\bibfnamefont {J.}~\bibnamefont {Meng}},\ }\href
  {https://doi.org/10.1103/PhysRevC.95.044301} {\bibfield  {journal} {\bibinfo
  {journal} {Phys. Rev. C}\ }\textbf {\bibinfo {volume} {95}},\ \bibinfo
  {pages} {044301} (\bibinfo {year} {2017})}\BibitemShut {NoStop}%
\bibitem [{\citenamefont {Wang}\ \emph {et~al.}(2020)\citenamefont {Wang},
  \citenamefont {Naito}, \citenamefont {Liang},\ and\ \citenamefont
  {Long}}]{Wang2020}%
  \BibitemOpen
  \bibfield  {author} {\bibinfo {author} {\bibfnamefont {Z.}~\bibnamefont
  {Wang}}, \bibinfo {author} {\bibfnamefont {T.}~\bibnamefont {Naito}},
  \bibinfo {author} {\bibfnamefont {H.}~\bibnamefont {Liang}},\ and\ \bibinfo
  {author} {\bibfnamefont {W.~H.}\ \bibnamefont {Long}},\ }\href
  {https://doi.org/10.1103/PhysRevC.101.064306} {\bibfield  {journal} {\bibinfo
   {journal} {Phys. Rev. C}\ }\textbf {\bibinfo {volume} {101}},\ \bibinfo
  {pages} {064306} (\bibinfo {year} {2020})}\BibitemShut {NoStop}%
\bibitem [{\citenamefont {Berger}\ \emph {et~al.}(1984)\citenamefont {Berger},
  \citenamefont {Girod},\ and\ \citenamefont {Gogny}}]{Berger1984}%
  \BibitemOpen
  \bibfield  {author} {\bibinfo {author} {\bibfnamefont {J.}~\bibnamefont
  {Berger}}, \bibinfo {author} {\bibfnamefont {M.}~\bibnamefont {Girod}},\ and\
  \bibinfo {author} {\bibfnamefont {D.}~\bibnamefont {Gogny}},\ }\href
  {https://doi.org/https://doi.org/10.1016/0375-9474(84)90240-9} {\bibfield
  {journal} {\bibinfo  {journal} {Nuclear Physics A}\ }\textbf {\bibinfo
  {volume} {428}},\ \bibinfo {pages} {23} (\bibinfo {year} {1984})}\BibitemShut
  {NoStop}%
\bibitem [{\citenamefont {Marketin}\ \emph {et~al.}(2007)\citenamefont
  {Marketin}, \citenamefont {Vretenar},\ and\ \citenamefont
  {Ring}}]{Marketin2007}%
  \BibitemOpen
  \bibfield  {author} {\bibinfo {author} {\bibfnamefont {T.}~\bibnamefont
  {Marketin}}, \bibinfo {author} {\bibfnamefont {D.}~\bibnamefont {Vretenar}},\
  and\ \bibinfo {author} {\bibfnamefont {P.}~\bibnamefont {Ring}},\ }\href
  {https://doi.org/10.1103/PhysRevC.75.024304} {\bibfield  {journal} {\bibinfo
  {journal} {Phys. Rev. C}\ }\textbf {\bibinfo {volume} {75}},\ \bibinfo
  {pages} {024304} (\bibinfo {year} {2007})}\BibitemShut {NoStop}%
\bibitem [{\citenamefont {Lalazissis}\ \emph {et~al.}(2005)\citenamefont
  {Lalazissis}, \citenamefont {Nik\ifmmode \check{s}\else
  \v{s}\fi{}i\ifmmode~\acute{c}\else \'{c}\fi{}}, \citenamefont {Vretenar},\
  and\ \citenamefont {Ring}}]{Lalazissis2005}%
  \BibitemOpen
  \bibfield  {author} {\bibinfo {author} {\bibfnamefont {G.~A.}\ \bibnamefont
  {Lalazissis}}, \bibinfo {author} {\bibfnamefont {T.}~\bibnamefont
  {Nik\ifmmode \check{s}\else \v{s}\fi{}i\ifmmode~\acute{c}\else \'{c}\fi{}}},
  \bibinfo {author} {\bibfnamefont {D.}~\bibnamefont {Vretenar}},\ and\
  \bibinfo {author} {\bibfnamefont {P.}~\bibnamefont {Ring}},\ }\href
  {https://doi.org/10.1103/PhysRevC.71.024312} {\bibfield  {journal} {\bibinfo
  {journal} {Phys. Rev. C}\ }\textbf {\bibinfo {volume} {71}},\ \bibinfo
  {pages} {024312} (\bibinfo {year} {2005})}\BibitemShut {NoStop}%
\bibitem [{\citenamefont {Kondev}\ \emph {et~al.}(2021)\citenamefont {Kondev},
  \citenamefont {Wang}, \citenamefont {Huang}, \citenamefont {Naimi},\ and\
  \citenamefont {Audi}}]{Kondev2021}%
  \BibitemOpen
  \bibfield  {author} {\bibinfo {author} {\bibfnamefont {F.}~\bibnamefont
  {Kondev}}, \bibinfo {author} {\bibfnamefont {M.}~\bibnamefont {Wang}},
  \bibinfo {author} {\bibfnamefont {W.}~\bibnamefont {Huang}}, \bibinfo
  {author} {\bibfnamefont {S.}~\bibnamefont {Naimi}},\ and\ \bibinfo {author}
  {\bibfnamefont {G.}~\bibnamefont {Audi}},\ }\href
  {https://doi.org/10.1088/1674-1137/abddae} {\bibfield  {journal} {\bibinfo
  {journal} {Chinese Physics C}\ }\textbf {\bibinfo {volume} {45}},\ \bibinfo
  {pages} {030001} (\bibinfo {year} {2021})}\BibitemShut {NoStop}%
\bibitem [{\citenamefont {M\"oller}\ \emph {et~al.}(2019)\citenamefont
  {M\"oller}, \citenamefont {Mumpower}, \citenamefont {Kawano},\ and\
  \citenamefont {Myers}}]{Moeller2019}%
  \BibitemOpen
  \bibfield  {author} {\bibinfo {author} {\bibfnamefont {P.}~\bibnamefont
  {M\"oller}}, \bibinfo {author} {\bibfnamefont {M.}~\bibnamefont {Mumpower}},
  \bibinfo {author} {\bibfnamefont {T.}~\bibnamefont {Kawano}},\ and\ \bibinfo
  {author} {\bibfnamefont {W.}~\bibnamefont {Myers}},\ }\href
  {https://doi.org/https://doi.org/10.1016/j.adt.2018.03.003} {\bibfield
  {journal} {\bibinfo  {journal} {At. Data Nucl. Data Tables}\ }\textbf
  {\bibinfo {volume} {125}},\ \bibinfo {pages} {1} (\bibinfo {year}
  {2019})}\BibitemShut {NoStop}%
\bibitem [{\citenamefont {Wang}\ \emph {et~al.}(2021)\citenamefont {Wang},
  \citenamefont {Huang}, \citenamefont {Kondev}, \citenamefont {Audi},\ and\
  \citenamefont {Naimi}}]{Wang2021}%
  \BibitemOpen
  \bibfield  {author} {\bibinfo {author} {\bibfnamefont {M.}~\bibnamefont
  {Wang}}, \bibinfo {author} {\bibfnamefont {W.}~\bibnamefont {Huang}},
  \bibinfo {author} {\bibfnamefont {F.}~\bibnamefont {Kondev}}, \bibinfo
  {author} {\bibfnamefont {G.}~\bibnamefont {Audi}},\ and\ \bibinfo {author}
  {\bibfnamefont {S.}~\bibnamefont {Naimi}},\ }\href
  {https://doi.org/10.1088/1674-1137/abddaf} {\bibfield  {journal} {\bibinfo
  {journal} {Chinese Physics C}\ }\textbf {\bibinfo {volume} {45}},\ \bibinfo
  {pages} {030003} (\bibinfo {year} {2021})}\BibitemShut {NoStop}%
\bibitem [{\citenamefont {Borzov}(2003)}]{Borzov2003}%
  \BibitemOpen
  \bibfield  {author} {\bibinfo {author} {\bibfnamefont {I.~N.}\ \bibnamefont
  {Borzov}},\ }\href {https://doi.org/10.1103/PhysRevC.67.025802} {\bibfield
  {journal} {\bibinfo  {journal} {Phys. Rev. C}\ }\textbf {\bibinfo {volume}
  {67}},\ \bibinfo {pages} {025802} (\bibinfo {year} {2003})}\BibitemShut
  {NoStop}%
\bibitem [{\citenamefont {Borzov}(2005)}]{Borzov2005}%
  \BibitemOpen
  \bibfield  {author} {\bibinfo {author} {\bibfnamefont {I.~N.}\ \bibnamefont
  {Borzov}},\ }\href {https://doi.org/10.1103/PhysRevC.71.065801} {\bibfield
  {journal} {\bibinfo  {journal} {Phys. Rev. C}\ }\textbf {\bibinfo {volume}
  {71}},\ \bibinfo {pages} {065801} (\bibinfo {year} {2005})}\BibitemShut
  {NoStop}%
\bibitem [{\citenamefont {Yoshida}\ \emph {et~al.}(2018)\citenamefont
  {Yoshida}, \citenamefont {Utsuno}, \citenamefont {Shimizu},\ and\
  \citenamefont {Otsuka}}]{Yoshida2018}%
  \BibitemOpen
  \bibfield  {author} {\bibinfo {author} {\bibfnamefont {S.}~\bibnamefont
  {Yoshida}}, \bibinfo {author} {\bibfnamefont {Y.}~\bibnamefont {Utsuno}},
  \bibinfo {author} {\bibfnamefont {N.}~\bibnamefont {Shimizu}},\ and\ \bibinfo
  {author} {\bibfnamefont {T.}~\bibnamefont {Otsuka}},\ }\href
  {https://doi.org/10.1103/PhysRevC.97.054321} {\bibfield  {journal} {\bibinfo
  {journal} {Phys. Rev. C}\ }\textbf {\bibinfo {volume} {97}},\ \bibinfo
  {pages} {054321} (\bibinfo {year} {2018})}\BibitemShut {NoStop}%
\end{thebibliography}%

\end{document}